\documentclass[printer]{aa}
\usepackage{natbib}
\usepackage[multi-part-units=single]{siunitx}
\DeclareSIUnit\angstrom{\text {Å}}
 \DeclareSIUnit\gauss{G}
 \DeclareSIUnit\kilogauss{kG}
 \DeclareSIUnit\maxwell{Mx}
\usepackage{amsmath}
\usepackage{nccmath}
\usepackage{graphicx }
\usepackage{xcolor}
\usepackage{soul}

\begin{document}
\title{The role of the chromospheric magnetic canopy in the formation of a sunspot penumbra}

\author{P. Lindner \inst{1}  
\and C. Kuckein  \inst{2,3} 
\and S.J. González Manrique \inst{1,2,3,4}
\and N. Bello Gonz\'alez \inst{1} 
\and L. Kleint  \inst{5,6} 
\and T. Berkefeld  \inst{1} }

\titlerunning{The role of the chromospheric magnetic canopy in penumbra formation}
\authorrunning{Lindner et al.} 

\institute{
Leibniz-Institut für Sonnenphysik (KIS), Schöneckstr. 6, 79104 Freiburg, Germany 
\and  
Instituto de Astrof\'{i}sica de Canarias (IAC), V\'{i}a L\'{a}ctea s/n, 38205 La Laguna Tenerife, Spain 
\and 
Departamento de Astrof\'{\i}sica, Universidad de La Laguna 38205, La Laguna, Tenerife, Spain 
\and  
Astronomical Institute, Slovak Academy of Sciences, 05960 Tatranská Lomnica, Slovak Republic
\and 
University of Geneva, CUI, 1227 Carouge, Switzerland
\and
Astronomical Institute, University of Bern, Sidlerstrasse 5, 3012 Bern, Switzerland
}

\abstract{While it is being conjectured that a chromospheric canopy plays a role in penumbra formation, it has been difficult to find observational evidence of the connectivity between the photosphere and the chromosphere.}
{We investigate the existence of a chromospheric canopy as a necessary condition for the formation of a penumbra and aim to find the origin of the inclined magnetic fields.}
{Spectropolarimetric observations of NOAA AR 12776 from the GRIS@GREGOR instrument were analyzed. Atmospheric parameters were obtained from the deep photospheric \ion{Ca}{i}~10839~\SI{}{\angstrom} line (VFISV inversion code), the mostly photospheric \ion{Si}{i}~10827~\SI{}{\angstrom} line (SIR inversion code) and the chromospheric \ion{He}{i}~10830~\SI{}{\angstrom} triplet (HAZEL inversion code). We compared the photospheric and chromospheric magnetic topology of a sunspot sector with a fully fledged penumbra to a sector where no penumbra formed. Additionally, imaging data from the BBI@GREGOR instrument in TiO-band and G-band were analyzed.}
{In the deepest atmospheric layers, we find that the magnetic properties (inclination and field strength distribution) measured on the sunspot sector with fully fledged penumbra are similar to those measured on the sector without penumbra. Yet, in higher layers, magnetic properties are different. In the region showing no penumbra, almost vertical chromospheric magnetic fields are observed. Additionally, thin filamentary structures with a maximum width of \SI{0.1}{\arcsec} are seen in photospheric high-resolution TiO-band images in this region.}
{The existence of a penumbra is found to be discriminated by the conditions in the chromosphere. This indicates that a chromospheric canopy is a necessary condition for the formation of a penumbra. However, our results demonstrate that inclined fields in the chromospheric canopy are not needed for the development of inclined fields in the photosphere. We question the `fallen-magnetic-flux-tubes' penumbra formation scenario and favor a scenario, in which inclined fields emerge from below the surface and are blocked by the overlying chromospheric canopy.}
\keywords{sunspots, Sun: photosphere, Sun: chromosphere, Sun: magnetic fields, Sun: evolution}

\maketitle

  %==============================================
  %==============================================

\section{Introduction}
Sunspots consist of an inner part, the umbra, which harbors mostly vertical and strong magnetic fields with field strength values of up to several kilogauss, and an outer part, the penumbra, where magnetic fields are inclined and weaker, although values of more than \SI{2000}{\gauss} can be observed \citep{2011LRSP....8....4B}. In the penumbra, a filamentary structure is generally observed in the photosphere. The magnetic field vector alternates between being stronger and less inclined in the spines and being weaker and more inclined in the intraspines, as shown by spectropolarimetric observations \citep[e.g.][]{2008ApJ...689L..69S,2013A&A...557A..25T,2016A&A...596A...2B}. This is in agreement with the model of the uncombed structure of the penumbra \citep{1993A&A...275..283S}, in which two separate magnetic components with different inclinations are interlaced in the penumbra. Stable sunspots have also been observed to possess, among others, the following spatially averaged properties: First, the outward directed Evershed flow \citep{1909MNRAS..69..454E} is observed in the penumbra. Azimuthally averaged, this flow is predominantly horizontal, with an up-flow component in the inner and a downflow component in the outer penumbra \citep{2000A&A...358.1122S}. Second, a chromospheric canopy in the form of horizontal magnetic fields is observed above the penumbra of sunspots \citep[e.g.][]{2017A&A...604A..98J}, while the imprints of the spine and intraspine structure are seen up to at least the middle chromosphere \citep{2019ApJ...873..126M}. Third, the so called inverse Evershed flow is observed in the chromosphere as radial in-flows (opposite to the outflowing photospheric Evershed flow) in the superpenumbra (the region above and beyond the photospheric penumbra). This has first been observed by \citet{1975SoPh...43...91M} and more recently investigated by, for example, \citet{2017A&A...604A..98J} and \citet{2020ApJ...891..119B}.

\noindent The formation of a penumbra, however, is still not well understood, mostly because high-resolution observations of forming penumbrae are still scarce. \citet{2010AN....331..563S, 2010A&A...512L...1S} and \citet{2012A&A...537A..19R} presented a \SI{4.5}{\hour} imaging and spectropolarimetric data set from a sunspot developing a penumbra initially on the side facing away from the Active Region (AR) opposite polarity. On the side facing the AR opposite polarity, flux emergence was observed in the form of bipoles, that partially merged with the spot. \citet{2013ApJ...771L...3R} and \citet{2016ApJ...825...75M,2017ApJ...834...76M}  also recorded spectropolarimetric data of a developing penumbra, and showed that the penumbra can initially also form on the spot side facing the opposite polarity. Additionally, they also observed an annular zone around the spot in the photosphere, in which inclined fields were present before the penumbra forms. \citet{2018ApJ...855...58M} showed proof that penumbra formation can start on both sides of a sunspot, based on Spaceweather HMI Active Region Patch (SHARP) data \citep{2014SoPh..289.3549B} from the Helioseismic and Magnetic Imager (HMI) \citep{2012SoPh..275..229S} onboard the Solar Dynamics Observatory (SDO) \citep{2012SoPh..275....3P}.

\noindent High-resolution spectropolarimetric data from the chromosphere during penumbra formation are still missing. In a literature search, we found that chromospheric fields during penumbra formation have, up to now, only been analyzed with high-resolution based on imaging or spectroscopic observations. Using \ion{Ca}{ii}~H images from SOT telescope \citep{2008SoPh..249..167T} onboard the Hinode space observatory \citep{2007SoPh..243....3K}, \citet{2012ApJ...747L..18S} observed an annular zone with reduced brightness in chromospheric images hours before the formation of penumbral filaments in the photosphere. This provided first observational evidence for a chromospheric canopy playing a major role during penumbra formation.  A special case of penumbral structures in between two sunspots in the decaying phase was analyzed by \citet{2021A&A...653A..93M}. Based on magnetic field information extrapolated from HMI data, the authors suggest that the penumbra formed only in the presence of an overlying canopy. Further evidence comes from numerical simulations, in which a sunspot develops a penumbra or not depending on the degree at which magnetic fields are forced to be inclined at the top of the simulation box \citep{2012ApJ...750...62R}. Other simulations show that the inclined fields of the canopy can be formed by flux emergence \citep{2016ApJ...831L...4M} and, later on, influence penumbra formation.

\noindent Concerning the interplay between photospheric and chromospheric magnetic fields during the formation of a penumbra around a sunspot, two scenarios have been competing over the last years: \\
\citet{2012ApJ...747L..18S}, \citet{2014ApJ...784...10R} and \citet{2016ApJ...825...75M} suggest that inclined magnetic fields in the photosphere can be formed by the transfer of inclined field lines from the chromosphere down to the photosphere. This scenario is often referred to as the "falling field lines scenario". The authors used this mechanism to explain photospheric redshifts around the protospot prior to the penumbra formation and also to explain a slightly more horizontal field (\SI{80}{\degree}) compared to a normal penumbra (\SI{70}{\degree}) at the beginning of penumbra formation. The scenario goes back to the `fallen magnetic flux tubes' model by \citet{1992ApJ...388..211W}. \\
In a second scenario, flux emergence happens in the photosphere and the inclined field lines cannot continue rising because they are trapped by the overlying canopy fields in the chromosphere. This mechanism is based on a scenario proposed by \citet{1998ApJ...507..454L} and was brought up again by \citet{2013ApJ...769L..18L}, who analyzed the formation of several penumbral features with both photospheric (Titanium Oxide band at \SI{7057}\angstrom) and chromospheric (H-alpha) images. Penumbrae formed in regions that showed both flux emergence and chromospheric threads in H-alpha images, which are interpreted as parts of the chromospheric canopy. The formation of the penumbra was then accompanied by new chromospheric threads with a slightly different orientation, aligned with the photospheric penumbra. Together with the observed flux emergence, this was interpreted as penumbral magnetic fields emerging from below rather than falling from the chromospheric canopy. Another argument supporting this scenario are observations by \citet{2018ApJ...857...21L,2019ApJ...886..149L}, who also showed that a penumbra can form in regions where flux emergence happened. \citet{2012A&A...539A.131K} and \citet{2014ApJ...787...57Z} also ascribed similar formation mechanisms to orphan penumbrae. \\
All of the observations favoring one or the other scenario, however, are missing spectropolarimetric data of both chromospheric and photospheric spectral lines and therefore information about the magnetic field vector. The new generation of solar instruments can provide such data, but the chances to observe a penumbra formation event are low. It has, therefore, still not been possible to discriminate between the different scenarios of penumbra formation. In this paper, we present spectropolarimetric observational data including both photospheric and chromospheric lines of a sunspot with a partial penumbra. We conclude that our results favor the second penumbra formation scenario in which inclined fields emerge from deep layers and are blocked from further rising by a chromospheric canopy.
  
\section{Data and calibration}
\label{data}
\begin{figure}[h]
\centering
	\includegraphics[width=8.8cm]{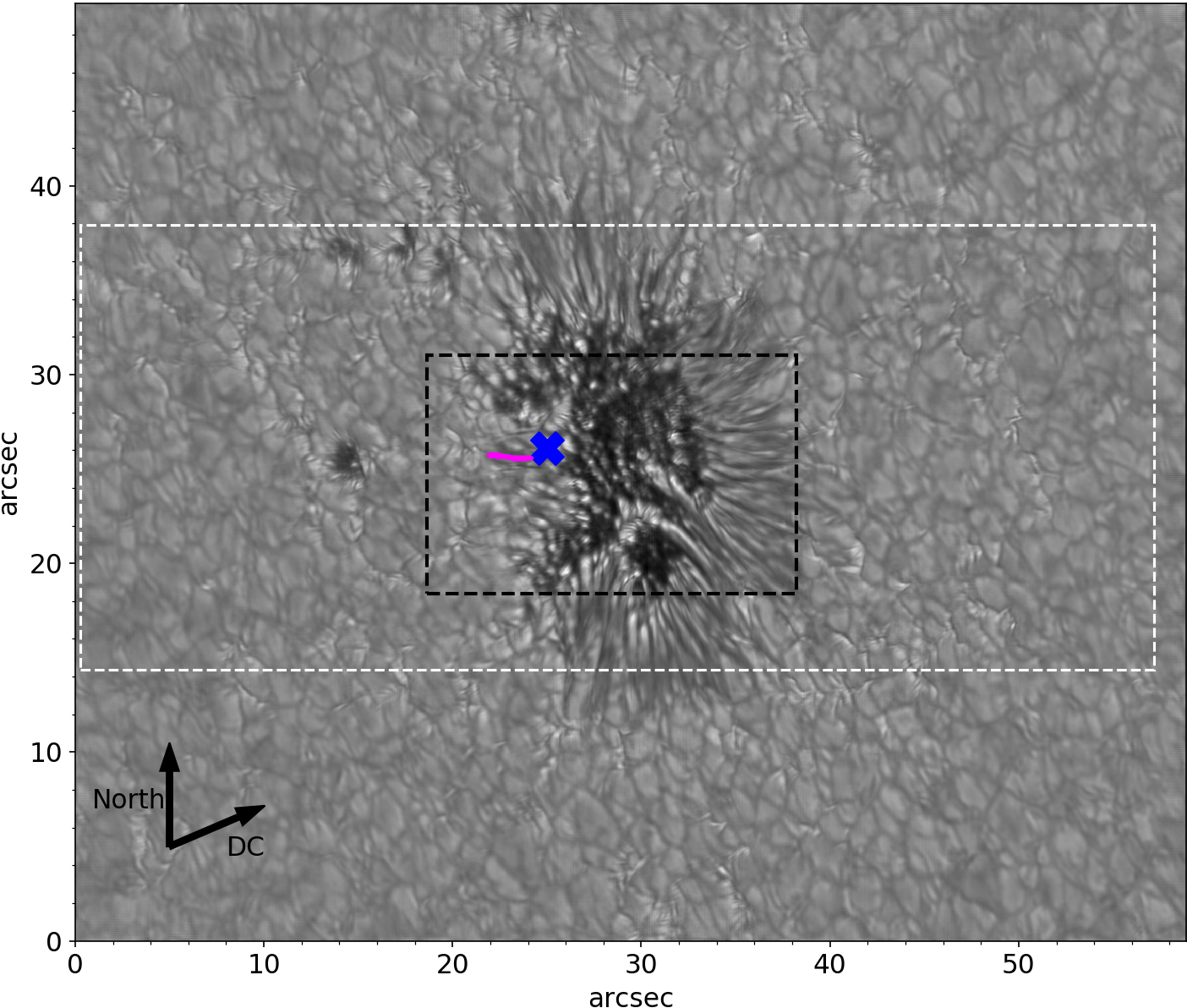}
    \caption{Speckle-reconstructed image with the TiO-band filter showing the sunspot with a partially developed penumbra. The contour of one thin bright filament (see Sect.~\ref{morph}) adjoining the left (eastern) side of the umbra is shown in magenta. The black-dashed box depicts the region shown in Fig.~\ref{inv_maps}.
    The heliocentric angle of this observations is $\approx$ \SI{55}{\degree}. The area scanned with GRIS is marked with a white-dashed box. The scanning direction was from bottom to top.} 
     \label{overview}
\end{figure}

Observations were conducted with the GREGOR telescope \citep{2012AN....333..796S,2012AN....333..863B}, during a Science Verification Phase that followed a complete redesign of the GREGOR optics laboratory \citep{2020A&A...641A..27K}. The target was a sunspot with the NOAA AR number 12776 showing a partial penumbra. During our observations, the AR was located south-east from solar disk center with a $\mu$ value (cosine of the heliocentric angle) of $0.639$. An image with the full field of view (FOV) obtained with the GREGOR Broad Band Imager (BBI) is shown in Fig.~\ref{overview}. In this work, we are showing data from 2020-10-16 between 08:12 UT and 09:08 UT. Until approximately 08:50 UT, the seeing conditions were stable with an average Fried parameter $r_0$ of \SI{9.1}{\cm}. After that, seeing conditions started to deteriorate.

\subsection{Imaging data}
Images were recorded using an adaptation of the BBI instrument \citep{2012AN....333..894V}. The two \emph{Andor Zyla} cameras were equipped with a G-band filter and an optimized titanium monoxide (TiO-band) filter centered at the TiO molecular band at \SI{7057}{\angstrom}. Observing in these molecular bands has proven to be successful for high-contrast imaging of the solar photosphere and sunspots \citep[see, e.g.][]{2003ApJ...589L.117B}. The cameras have 2560 x 2160 pixels and were set up with a pixel scale of 0.023 arcseconds per pixel. The telescope diffraction limit $ \frac{\lambda}{D}$ is \SI{0.10}{\arcsec} for the TiO-band wavelength and 0.06\arcsec\ for G-band. Exposure times were \SI{1.2}{\milli \second} for the TiO-band images and \SI{3.0}{\milli \second} and \SI{2.5}{\milli \second} for the G-band images. 
The BBI data were calibrated with a standard pipeline, including a dark image correction and a flat field correction. The images were then reconstructed using the KISIP speckle code \citep{2008A&A...488..375W}, however, without taking into account the central obscuration of GREGOR, leading to small variations in contrast from the ground truth. After the reconstruction, a Butterworth filter was applied to all images to remove a pattern of vertical stripes arising from the camera. Finally, the G-band images were aligned to the respective temporally closest TiO-band image using a translational shift.

%(1083e-9/1.4)*206264.8

\subsection{GRIS spectropolarimetric data}
\label{grisdata}
Spectropolarimetric data were obtained with the GRegor Infrared Spectrograph \citep[GRIS,][]{2012AN....333..872C}. Three scans were performed with 180 scan positions for each map. A step size of \SI{0.135}{\arcsec} was chosen, so that the resulting slit-reconstructed maps have the same image scale in x and y direction. The first and, partially, the second scan were acquired under good seeing conditions and the third scan with only mediocre seeing conditions. The exposure time was \SI{100}{\milli \second} with 10 accumulations, leading to one scan taking roughly \SI{18}{\minute}. The spectral range was chosen to be around the \ion{He}{i}~10830~\SI{}{\angstrom}~triplet. In addition to this helium triplet, the spectral window included the \ion{Si}{i}~10827~\SI{}{\angstrom} and the \ion{Ca}{i}~10839~\SI{}{\angstrom} line. GRIS level1 data were created using the gris\_v8 pipeline by M. Collados (IAC, private communication). This version corrects for a sub-pixel misalignment between the two beams. A slit-reconstructed continuum map of the first GRIS scan, cropped around the sunspot, is shown in Fig.~\ref{inv_maps}. An absolute wavelength calibration was carried out taking into account Earth's orbital motion, solar and terrestrial rotation, as well as the gravity redshift \citep{1997ApJ...474..810M,2012A&A...542A.112K}, using telluric lines observed simultaneously within the spectral window.

\subsection{HMI data}
We utilized the ``sharp\_cea'' HMI data series \citep{2014SoPh..289.3549B} that has a cadence of 720 seconds and offers de-projected atmospheric data. HMI data has a spatial resolution of \SI{1}{\arcsec} \citep{2014SoPh..289.3483H}. Data access and handling was done using the SunPy open source software package \citep{sunpy_community2020}.

\section{Inversions}
\begin{figure*}[h]
\centering
   \includegraphics[width=17cm]{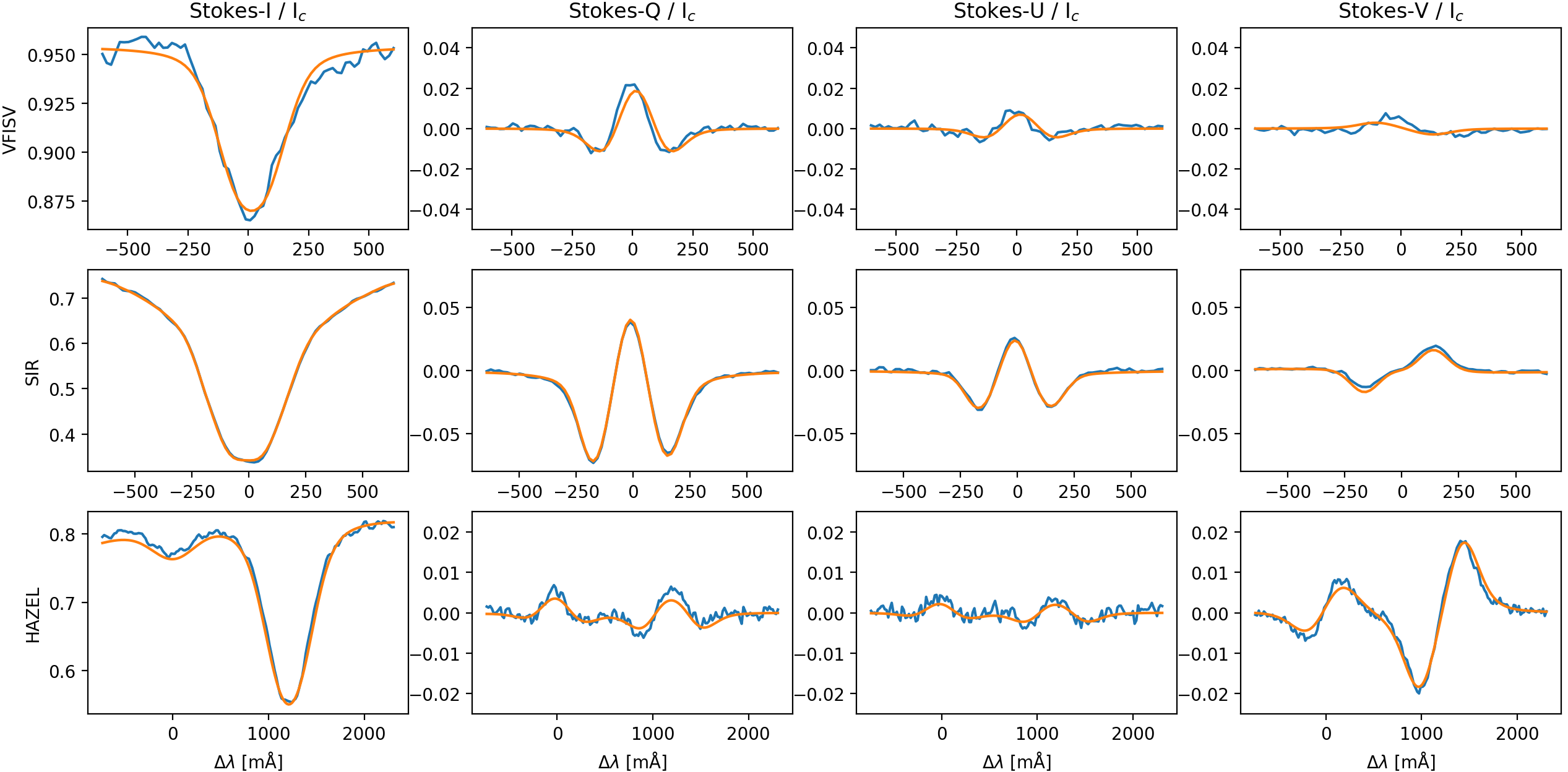}
     \caption{Spectral profiles (observations  in blue, inversion fit in orange) of a pixel from the region showing TBFs. The location of this pixel is depicted in Fig.~\ref{overview} with a blue cross. Top row: \ion{Ca}{I}~10839~\SI{}{\angstrom} line, central row: \ion{Si}{I}~10827~\SI{}{\angstrom} line, bottom row: \ion{He}{I}~10830~\SI{}{\angstrom} triplet. All spectra are normalized to the quiet-sun continuum value. The spectra for pixels from other regions are shown in Appendix~\ref{app_examspecs}.}
     \label{examspec}
\end{figure*}

The slit-reconstructed GRIS maps include spectropolarimetric data. Inversion tools can be applied to obtain atmospheric parameters from the full-Stokes spectra. Due to the larger infrared wavelength, both the spatial resolution and the contrast is lower in the GRIS maps than in the BBI imaging data. The pixel size is \SI{0.135}{\arcsec}, so the thin bright filaments (TBFs) characterized in Section \ref{morph} based on the BBI imaging data are not resolved. Inversions were carried out to investigate why no penumbra formed in the eastern part of the sunspot and to characterize the atmospheric surroundings, in which the TBFs formed. Depending on the line formation characteristics, three different inversion codes were used: (1) The computationally fast VFISV inversion code for the deep photospheric \ion{Ca}{I}~10839~\SI{}{\angstrom} line, (2) the SIR code for the (mostly) photospheric \ion{Si}{i}~10827~\SI{}{\angstrom} line to obtain height dependent information, and (3) the HAZEL inversion code for the chromospheric \ion{He}{I}~10830~\SI{}{\angstrom} triplet, which takes Zeeman and Hanle effects into account. Each line was inverted with a separate model atmosphere. This allowed us to do a velocity calibration based on the resulting inversion maps separately for each line (see Sect.~\ref{velcal}). Figure \ref{examspec} shows fitted spectra of one example pixel at [\SI{23}{\arcsec}, \SI{23.5}{\arcsec}]. These coordinates refer to BBI full FOV coordinates as shown, for example, in Fig.~\ref{overview}).

\subsection{VFISV inversions}
The \ion{Ca}{i}~10839~\SI{}{\angstrom} line is a magnetically sensitive line ($g_{\rm eff} = 1.5$) that forms in the deep photosphere. \citet{2018A&A...617A..39F} obtained a mean formation height of \SI{64}{\kilo \meter}. As it has a narrow formation height and forms in the deep photosphere, we used the computationally fast Milne-Eddington inversion code VFISV \citep{2011SoPh..273..267B} to obtain atmospheric parameter by making use of the {\it GRISinv}\footnote{https://gitlab.leibniz-kis.de/sdc/gris/grisinv} tool by Vigeesh et al. (KIS) optimised for VFISV inversions of GRIS data.

\subsection{SIR inversions}
\label{sirinv}
The \ion{Si}{I}~10827~\SI{}{\angstrom} line is sensitive to atmospheric parameters in the upper photosphere. While the wings are formed under local thermodynamical equilibrium (LTE), only the central core shows effects of non-LTE formation mechanisms \citep{2008ApJ...682.1376B}. However, as shown by, for example, \citet{2012A&A...539A.131K} and \citet{2019A&A...630A.139K}, depth-dependent inversions can still be used to obtain atmospheric parameters under the LTE assumption, neglecting the smallest optical depths, to which the line core are sensitive to. The spectropolarimetric data of the \ion{Si}{I}~10827~\SI{}{\angstrom} line from the GRIS instrument was inverted with the SIR inversion code \citep{1992ApJ...398..375R}. A python-based wrapper written by Ricardo Gafeira\footnote{{https://gitlab.com/gafeira/parallel\_desire\_sir\_rh}} was used to initialize the inversion and to run it in parallel on multiple CPUs. Each pixel was inverted eight times with different initial-model atmospheres. After these eight runs, the one with the lowest $\chi ^2$ value (describing the squared difference between the observed spectral profile and the inversion fit) was chosen for each pixel. For each individual run, we used three cycles with (2, 3, 4) nodes in temperature, (1, 2, 2) nodes in the magnetic field strength, (1, 2, 3) nodes in the velocity, (1, 2, 2) nodes in the inclination angle of the magnetic field, and (1, 1, 1) node in the azimuth angle of the magnetic field. The built-in spatial straylight correction method from SIR was used with a quiet sun profile obtained from averaging over a quiet region on the western side of the FOV and an $\alpha$ value (setting the assumed relative amount of straylight) of $0.1$.

\subsection{HAZEL inversions}
\label{hazel_inv}
In order to obtain atmospheric parameters of the chromosphere, the \ion{He}{I}~10830~\SI{}{\angstrom}~triplet (observed with GRIS) was analyzed. Single-atmospheric-component inversions of the Stokes profiles were performed with the HAZEL inversion code \citep{2008ApJ...683..542A}, assuming magnetic field parameters and the velocity to be constant in height within the slab. The HAZEL code takes Zeeman and Hanle effects into account. We used three cycles, of which the first one was only used to obtain the optical depth and the velocity from Stokes-$I$. In the second and third cycle, polarimetric spectra were added and the magnetic field vector was obtained. We included weights for Stokes-$Q$ and Stokes-$U$ twice the weight of Stokes-$V$. A straylight correction was performed in the same way as for the SIR inversion (see Sect. \ref{sirinv}). 

\noindent Using this inversion setup, the inversion code produced good fits to the Stokes profiles for most of the pixels. However, in a specific region in which the circular polarization signal was low and the linear polarization signal was high, we were not able to find an inversion setup that produced good fits for all Stokes parameters at the same time.  We do not have a clear explanation for this behavior, but have found similar problems in other publications: For example, it is possible that the Stokes-$V$ signal forms in a different atmospheric height than Stokes-$Q$ and Stokes-$U$, as suggested by \citet{2019A&A...625A.128D}, or there may be 3D effects affecting the local anisotropy. \citet{2019A&A...625A.129D} showed that a two-component inversion may be required. The authors also showed, however, that this introduces a strong degeneracy of the solutions which made it impossible to infer atmospheric parameters. 
For this work, we decided to identify the problematic pixels and exclude them from our analysis of the magnetic field. Our best inversion run produced good fits for Stokes-$I$ for the full FOV and good fits for Stokes-$Q$, Stokes-$U$ and Stokes-$V$ outside the problematic region. In the problematic region, the inversion fit deviated strongly from the observed spectra. We were able to identify the problematic pixels based on a $\chi^2$ value, that only takes the difference between the fit spectrum and the observed spectrum for linear polarization into account:
\begin{align*}
\chi_{\mathrm{lin}}^2 = \frac{1}{N}  \left(\sum_{\lambda} (Q_{ \mathrm{obs}}(\lambda) - Q_{ \mathrm {fit}}(\lambda))^2  + (U_{ \mathrm{obs}}(\lambda) - U_{ \mathrm{fit}}(\lambda))^2 \right)
\end{align*}
where $N$ stands for the number of wavelength points. For all pixels with a $\chi_{\mathrm{lin}}^2$ value above $1.1 \cdot 10^{-5}$, the respective pixels were masked out in the magnetic field vector maps. We manually verified that the remaining pixels indeed show a proper fit and can therefore be used for an analysis of the magnetic field.

\begin{figure}[h]
\centering
	\includegraphics[width=8.8cm]{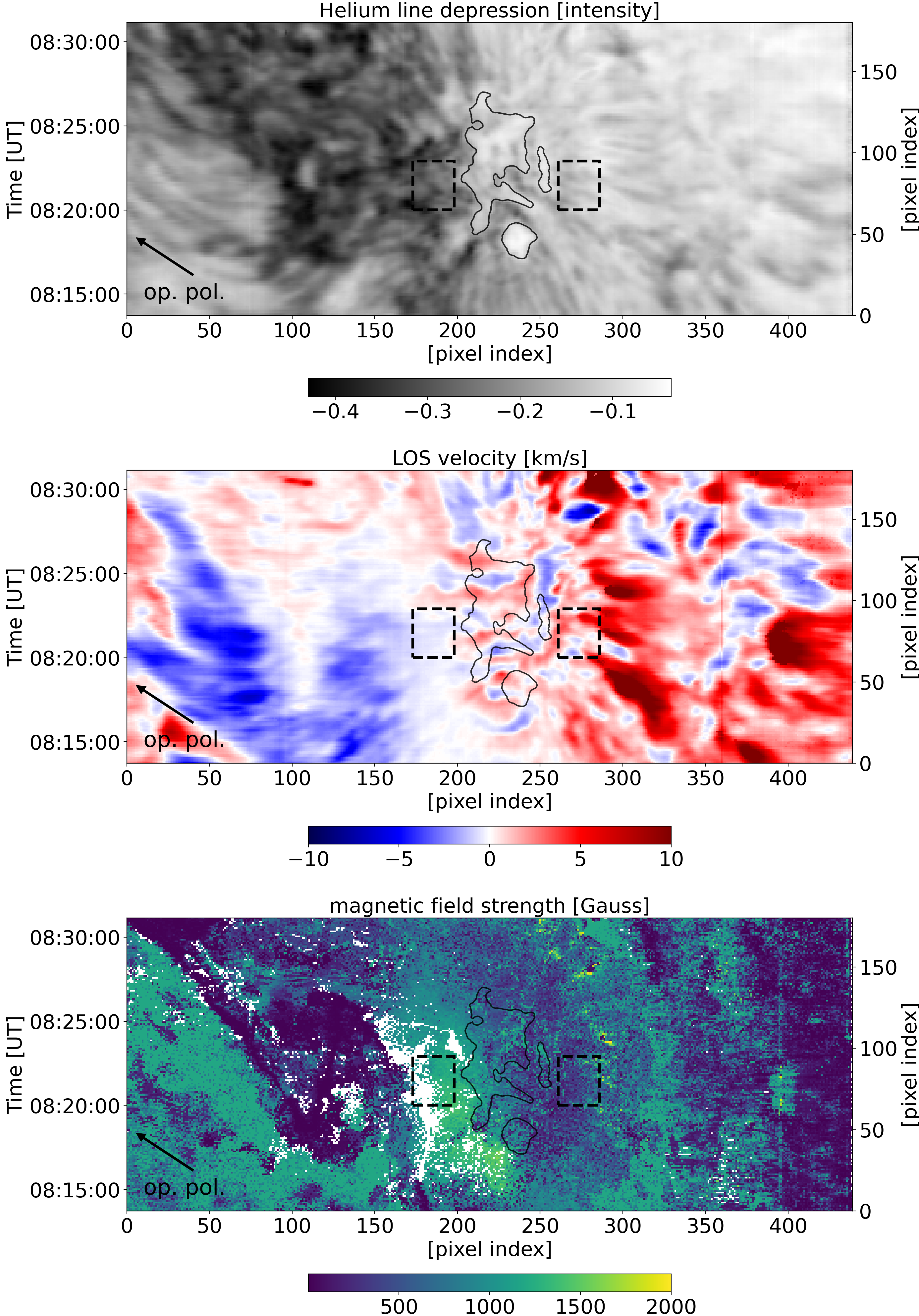}
     \caption{Top: Helium line depression map. The intensity is normalized to the quiet sun continuum value. Center: Line-of-sight velocity from the HAZEL inversions. Bottom: Total magnetic field strength from the HAZEL inversions with problematic pixels masked out in white (see Sect.~\ref{hazel_inv}). The black dotted boxes are analyzed in Sect.~\ref{compregions}. Coordinates refer to GRIS pixel numbers on the right and bottom axis and to time on the left axis. The full FOV is shown. The black arrow points towards the opposite polarity of the AR. Umbral contours from the continuum map are plotted in black.} 
     \label{hazel}
\end{figure}

\noindent Maps showing the velocity and the magnetic field strength over the full GRIS FOV are shown in Fig.~\ref{hazel}. The masked pixels are shown in white on the magnetic field strength map. Additionally, for each pixel, the line depression of the red component of the helium triplet is calculated by searching the intensity minimum of the helium line and subtracting the continuum value of the respective pixel. This map is also shown in Fig.~\ref{hazel}.

\subsection{Azimuthal disambiguation}
\label{grazam}
The \SI{180}{\degree} ambiguity of the azimuth angle of the magnetic field was resolved using the GRAZAM tool, which will be described in more detail in a future publication by \citet{grazam_inprep}. This tool automatically processes the inversion maps (e.g. correcting the orientation of the maps, etc.) and writes the input data for the AZAM code \citep{1995ApJ...446..877L}, which includes tools to manually edit azimuth maps with the computer mouse. Initially, a radially symmetric distribution of the magnetic field, centered in the sunspot umbra, was assumed. The remaining discontinuities in the maps were resolved by hand. This procedure was carried out separately for the VFISV map, the HAZEL map, and the SIR maps at different optical depths. On the disk center side of the sunspot (top right in the image plane), the magnetic field is parallel to the line-of-sight (LOS) in some pixels of the penumbra. Linear polarization signals are, therefore, small and the azimuth angle cannot be estimated well during the inversions. This uncertainty propagates through the transformation from the LOS reference to the local-reference-frame (LRF) and makes the disambiguation difficult in this region. In the inclination map in Fig.~\ref{inv_maps} from the VFISV inversion, a spurious discontinuity (sharp jump from dark blue to green-yellow) is visible in the top right of the image. These discontinuity lines are also present in HMI data (not shown here).

\subsection{Velocity calibration}
\label{velcal}
Although we used a carefully inferred absolute wavelength array (see Sect. \ref{grisdata}) as an input for the inversions, the velocity maps resulting from both the SIR and the VFISV inversions, initially showed an offset: The average over quiet sun pixels did not match the convective blueshift which is expected for a solar $\mu$ value (cosine of the heliocentric angle) of $0.639$, at which the sunspot was observed. We corrected the velocity offset by adding a constant value to the inversion maps for all $\log \tau$ values from the SIR results and for the one map of the VFISV result. This velocity value was chosen such that the quiet sun average velocity matches the reference value $v_{\textrm{ref}}$ that was measured with highest accuracy for several solar lines by \citet{2019A&A...624A..57L}. As they did not measure the \ion{Si}{I}~10827~\SI{}{\angstrom} line, we have chosen the \ion{Fe}{I}~5434~\SI{}{\angstrom} as a reference. This line has a formation height of \SI{550}{\kilo \meter} \citep[Table 1]{2019A&A...624A..57L}, which is similar to the formation height of the \ion{Si}{I}~10827~\SI{}{\angstrom} line core \citep[\SI{541}{\kilo \meter}, as calculated by][]{2008ApJ...682.1376B}. For the spectral resolution of $100\,000$ and a heliocentric position $\mu = 0.6$, a convective blueshift $v_{\textrm{ref}} = \SI{-6}{\meter / \second}$ is given by \citet[Table A.1]{2019A&A...624A..57L}. We have chosen our constant such, that the velocity map at $\log \tau = -2.7$ matches this value, which corresponds to the height where the maximum of the velocity response function is located in the quiet sun for this line \citep{2018A&A...617A..39F,2020A&A...634A..19G}. For the less frequently studied \ion{Ca}{I}~10839~\SI{}{\angstrom} line, \citet{2018A&A...617A..39F} obtained a mean formation height of \SI{64}{\kilo \meter}.  \citet{2019A&A...624A..57L} did not measure a line that matches this value, but the \ion{C}{I}~5381~\SI{}{\angstrom} and the \ion{Fe}{II}~6149~\SI{}{\angstrom} lines with formation heights of \SI{40}{\kilo \meter} and \SI{130}{\kilo \meter} (blue shift values of \SI{-821}{\meter / \second} and \SI{-404}{\meter / \second}, respectively), are close. We interpolated and used the value $v_{\textrm{ref}} =  \SI{-710}{\meter / \second}$ for the  \ion{Ca}{I}~10839~\SI{}{\angstrom} line. \\
After applying our correction, the average velocities in the umbra were below \SI{100}{\meter / \second} at $\log \tau = -2.0$ for the SIR inversions of the \ion{Si}{I}~10827~\SI{}{\angstrom} line and below \SI{50}{\meter / \second} for the VFISV inversions of the photospheric \ion{Ca}{I}~10839~\SI{}{\angstrom} line. This is in line with the expectation of significantly suppressed convection in the photospheric umbra \citep[see, e.g.][]{2018A&A...617A..19L} and we therefore regard our correction as consistent and successful.

\section{Analysis and results}

\subsection{Long-term flux evolution}
\begin{figure}[h!]
\centering
	\includegraphics[width=8.8cm]{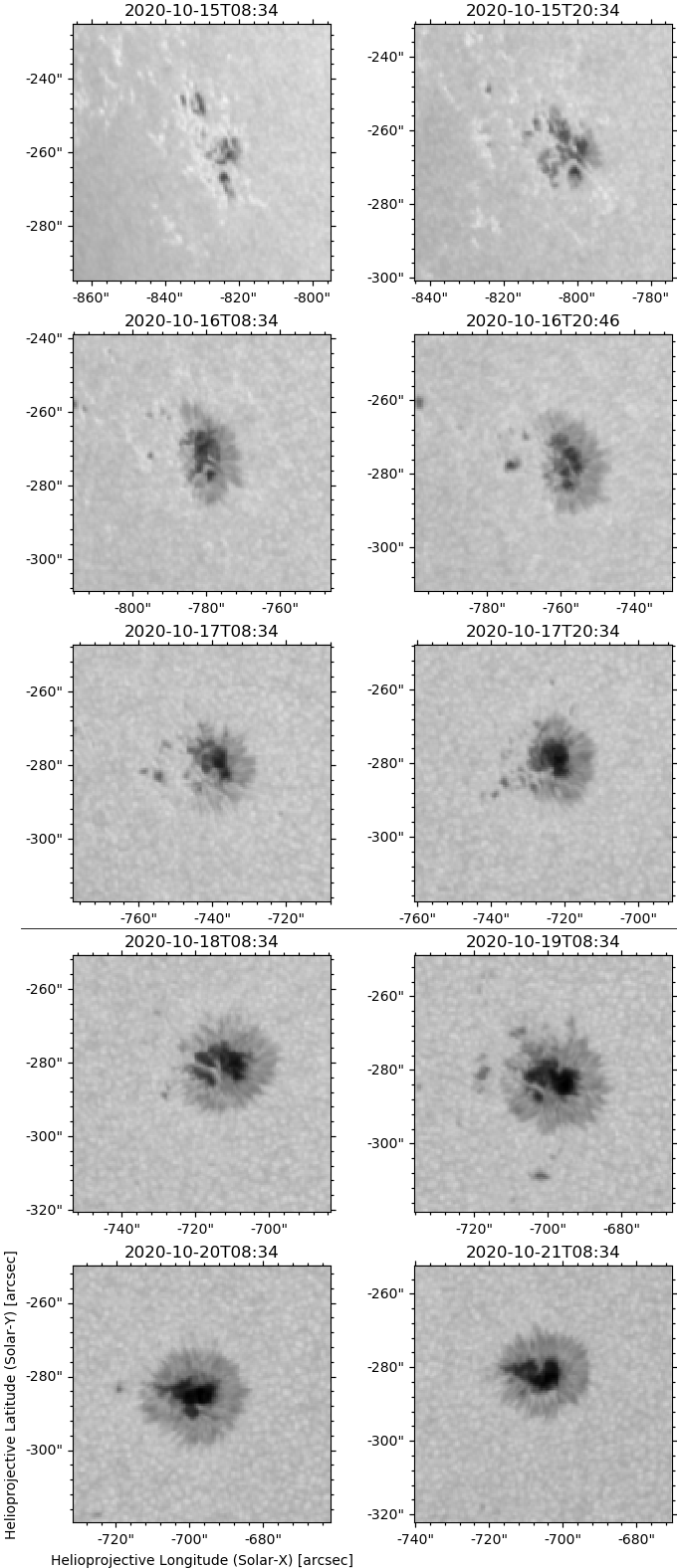}
     \caption{Continuum images from HMI showing the evolution of the sunspot in NOAA AR 12776 spanning around our ground-based observations on 2020-10-16 that started at 8:10 UT. The center right plot shows data from \SI{12}{\minute} after 2020-10-16T20:34, because that data was affected by a lunar eclipse ending at 19:53 UT. The time separation is \SI{12}{\hour} for the maps above the horizontal line and \SI{24}{\hour} for the maps below.} 
     \label{hmi_cont}
\end{figure}

We analyzed the evolution of the sunspot from 2020-10-15 to 2020-10-18 with HMI data. As shown in the continuum maps in Fig.~\ref{hmi_cont}, the sunspot formed at the beginning of this time range. On its western (right) side, the penumbra remained stable after the formation. On its eastern side, small pores or parts of the umbra kept changing shape, but no stable penumbra formed. Additional data from multiple days later (shown below the horizontal line in Fig.~\ref{hmi_cont}) yielded that, from 2020-10-20 (four days after our observations) onward, the sunspot had a penumbra that almost surrounded the whole umbra, but still left a small gap on the eastern side.

\begin{figure}[h!]
\centering
	\includegraphics[width=8.8cm]{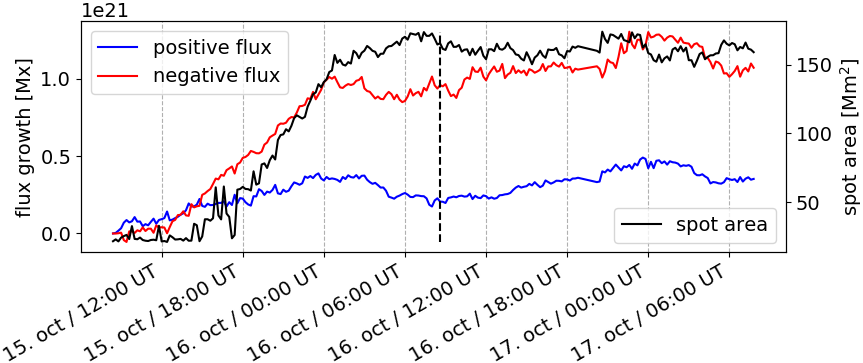}
     \caption{Temporal evolution of the positive and negative magnetic flux and spot size over two days, as calculated from HMI data. The vertical black-dashed line indicates the time of the GREGOR observations. Times are given in UT.} 
     \label{fluxevol}
\end{figure}

\noindent  In order to assess how much flux emergence is taking place around the sunspot during our observations, we calculated the temporal evolution of the magnetic flux from \SI{24}{\hour} before to \SI{24}{\hour} after our ground-based observations. We chose a rectangular FOV that covered the whole spot and its magnetic surroundings for this time interval. The absolute value of negative and positive fluxes were calculated separately by multiplying the value of the vertical component of the magnetic field with the area corresponding to one pixel and summing over all pixels with a negative or positive value. This is shown in Fig.~\ref{fluxevol}, with the positive and negative flux curves shifted in $y$-direction such that they start from zero. 

\noindent In addition, we made a simple estimation of the spot size by calculating the area that is covered by a contour in the HMI continuum image at an intensity value of 0.9. We selected the largest contour in our chosen FOV and manually verified that this contour indeed surrounded the sunspot. The evolution of the spot size is also shown in Fig.~\ref{fluxevol} the right-hand-side ordinate.

\noindent As shown in Fig.~\ref{fluxevol}, both the spot size and the negative flux increased until the early morning of 2021-10-16. Afterwards, both quantities reached a plateau. The amount of positive flux was lower than the negative flux and also showed no strong dynamics. The spot itself had a negative polarity.

\subsection{Morphological characterization of thin bright filaments}
\label{morph}
\begin{figure}[h]
\centering
	\includegraphics[width=8.8cm]{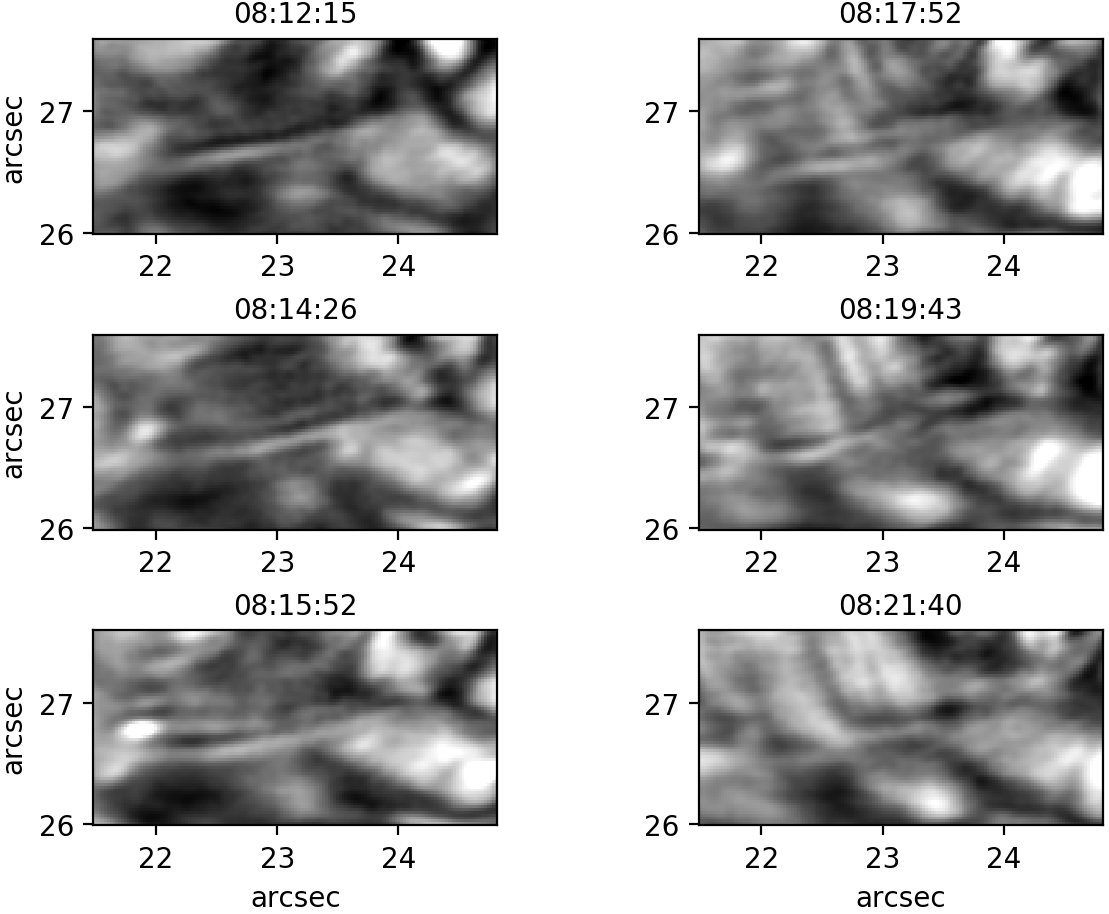}
     \caption{Evolution of one of the TBFs shown by images obtained with the TiO-band filter. Times are given in Universal Time. The spatial coordinates are given in arcseconds with respect to the full FOV of BBI. The location of this TBF is shown in Fig.~\ref{inv_maps} as one of the orange dotted lines.} 
     \label{timeevol}
\end{figure}

\begin{figure}[h]
\centering
	\includegraphics[width=8.8cm]{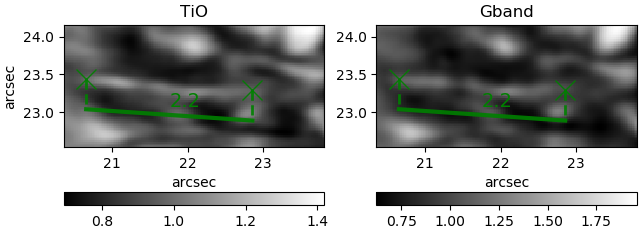}
     \caption{Example image of a thin bright filament obtained with a TiO-band (left) and a G-band (right) filter. The timing difference between the respective TiO-band and the G-band burst was \SI{349}{\milli \second}. The spatial coordinates are given in arcseconds with respect to the full FOV of BBI.} 
     \label{morph_example}
\end{figure}

In our GREGOR observations, no penumbra was visible on the left (East) side of the sunspot. The BBI images reveal that granulation was present with some of the granules being slightly elongated. In addition, on the same side, thin bright filaments (hereinafter referred to as TBFs) were seen that are mostly orientated east-west, that is, radially outward from the umbra. They are remarkably straight, almost no curvature is observed, and they seem to span multiple granules in some cases. This can be seen in the time series shown in Fig.~\ref{timeevol}. These TBFs can be clearly recognized and create the visual impression of forming separately from the granulation. Although they are also recognizable in the G-band images, they are better seen in TiO-band. Fig.~\ref{morph_example} shows an example of a TBF in G-band and TiO-band with a ruler that shows a length value in arcseconds.

\newpage % avoid for the itemize items to be split between three different columnns

\noindent Based on a visual inspection of the TiO-band images, 9 TBFs were identified, that clearly stood out from the granulation. Images of all of those TBFs can be found in Appendix~\ref{app1}.  Their width, length, and lifetimes were estimated manually and the following numbers were obtained: 

\begin{itemize}
\item Length: between \SI{0.8}{\arcsec} and \SI{2.2}{\arcsec} (average \SI{1.4}{\arcsec})
\item Width: seen as roughly \SI{0.1}{\arcsec}. This is at the telescope diffraction limit $ \frac{\lambda}{D}$ of \SI{0.10}{\arcsec} for TiO-band.
\item Lifetime: between \SI{2}{\minute} and \SI{10}{\minute} (average \SI{4}{\minute})
\end{itemize}

\noindent In many cases, it was not possible to unambiguously identify the time step, at which a TBF forms or decays. Similarly, the determination of the length includes a subjective choice of the endpoints of the TBF. Therefore, these numbers can serve as an orientation, but not as a highly accurate reference. Other characteristics we noticed are: 

\begin{itemize}
\item No inner structures is seen within the TBFs at this resolution.
\item The TBFs appear as a single entity and do not exist in bundles.
\item In 5 of the 9 events, a bright point-like feature is forming or moving close to one end of the filament. They resemble penumbral grains. In 4 out of these 5 cases, the bright point was located at the eastern end of the TBF and moved towards east (away from the umbra). The bright point-like features were better visible in the TiO-band images than in the G-band images.
\item The maximum brightness values (in TiO-band) inside the TBFs varies between 1.0 and 1.4 (average:~1.17) of the average quiet sun value.
\item TBFs were observed only on 2020-10-16. In imaging data from 2020-10-17 (also available, but not shown in this work), no TBFs were unambiguously identified. Only one bright elongated feature with a length of less than  \SI{1}{\arcsec} was observed on that date, that could also be part of an abnormal granulation.
\end{itemize}
In a literature search, we found publications where structures similar to the TBFs can be seen in the FOV of solar images \citep[e.g.][Fig.~2a]{2004ApJ...604..906R} and even in simulations \citep[Fig.~2]{2018ApJ...859L..26M}. They are, however, not discussed by the authors. Only the report by \citet{2012arXiv1207.6418Y} describes a `small scale filamentary flux emergence' event seen in TiO-band images which, in one image of a time sequence, looks similar to the TBFs described here. The length is stated to be approximately \SI{1}{\mega \meter} and it also had a bright dot appearing on one end. Close to this structure, a patch of horizontal magnetic field is observed in co-aligned HMI data. In the later stage of development, a bundle of multiple filaments appear in this location. Such bundles have also been observed by others \citep{2011ApJ...740...82L}, but are usually confined to (sometimes abnormal) granules. The TBFs in our case however, never appeared in bundles and were not confined to granular structures.

\subsection{Magnetic topology throughout the atmosphere}
\label{topology}
\begin{figure*}[h!]
\centering
   \includegraphics[width=17.4cm]{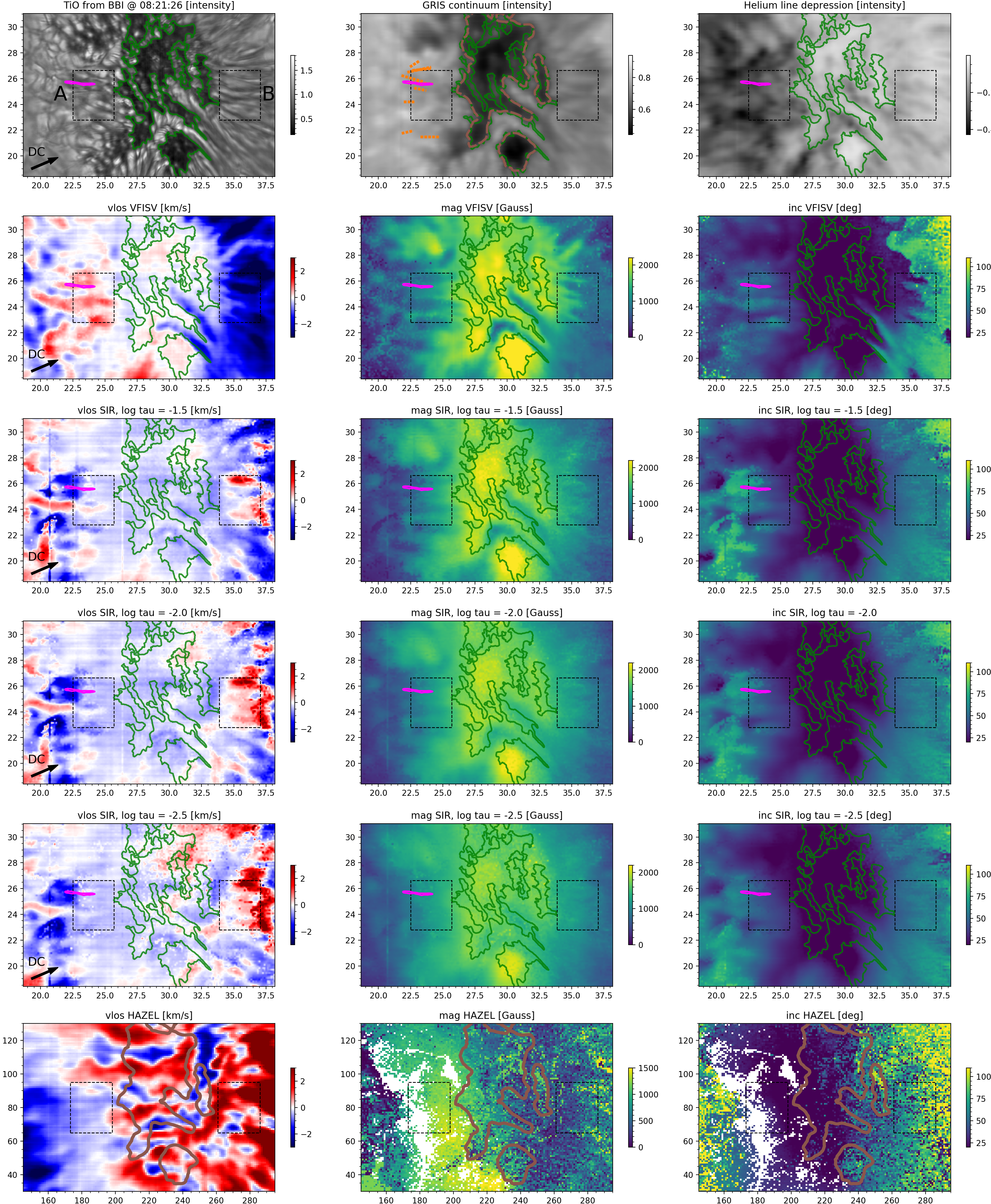}
     \caption{Maps of different atmospheric parameters produced by the inversions, cropped to the central region of the sunspot. Inclination angles refer to the LRF. Intensities are normalized by the average quiet sun value. A contour from the co-temporal TBF seen in the TiO-band images is plotted in magenta on all other maps, except for the HAZEL maps. The orange dotted lines on the GRIS continuum map represent the positions of the none-co-temporal TiO-band TBFs. Except for the last row, all maps were aligned to the BBI image shown in the top left and the coordinates refer to arcseconds. For the HAZEL results, no alignment was done to avoid an interpolation of masked (white) pixels. Therefore, coordinates refer to GRIS pixel numbers for the last row.}
     \label{inv_maps}
\end{figure*}

Using the inversion results from VFISV, SIR and HAZEL, the topology of the magnetic field vector and the LOS velocity can be analyzed not only in two dimensions, but also with respect to the atmospheric height. In Fig.~\ref{inv_maps}, cropped maps of the LOS velocity, the  magnetic field strength and its inclination angle in the LRF of all three inversion codes are shown. For the depth-dependent model from the SIR inversions, maps at optical depths of $\log \tau = (-1.5, -2.0, -2.5) $ are shown. In the magnetic field parameter maps from the HAZEL inversion, the white pixels are neglected, as described in Sect.~\ref{hazel_inv}. As expected, all maps show that the magnetic field is mostly vertical to the solar surface in the umbra, together with high values of the magnetic field strength. In the deepest atmospheric layer (VFISV maps), the field strength values are the highest with umbral values of more than \SI{2000}{\gauss}, which is in line with previous observations \citep[see][and references therein]{2018SoPh..293..120B}. In the penumbra, the well-known filamentary pattern in field strength and inclination maps is seen clearly in the VFISV maps. In the SIR maps, this pattern is weaker and fades towards higher layers. It is remarkable, however, that filamentary structures are also present on the eastern side (in and around region A) of the sunspot, where no penumbra is seen in continuum, TiO-band, and G-band images. Like the penumbra on the western side (in and around region B), which shows a classical spine and instraspine pattern, the eastern side (region A) also shows elongated structures of inclined magnetic fields and low field strength alternating with elongated structures of more vertical fields and high field strength values. The filamentation on both sides of the umbra is best visible in the deep photospheric inclination and field strength maps from VFISV. Without the knowledge of the continuum image, it would be difficult to tell whether a penumbra is present in and around region A or not. A quantitative comparison between the magnetic field properties of region A and region B follows in Sect.~\ref{compregions}. \\ The similarities between the eastern and the western side seem to be most prominent in lower layers. In the chromospheric HAZEL inversion maps, a clear difference between the eastern and the western region can be seen by a visual inspection of the cropped maps shown in Fig.~\ref{inv_maps}. The magnetic field strength is higher on the eastern side of the sunspot and the field is more vertically directed than on the western side. The field strength in the eastern side is also higher than in the umbral part of the spot. Interestingly, the region with high field strength values and almost vertically directed fields is at the location where the dark structure seen in the line depression map meets the umbral contour (Fig.~\ref{hazel}). In the maps showing the full FOV in Fig.~\ref{hazel}, one can see that the dark structure in the line depression map covers a large region at the north-eastern side of the FOV. It probably extends beyond the FOV, towards the direction of the opposite polarity of the AR.

\noindent After spatially co-aligning the BBI images with the inversion maps from the GRIS data, we were able to draw contours of a TBF (observed in TiO-band, see Sect.~\ref{morph}) onto the GRIS inversion maps. For one of those TBFs, the GRIS slit was at the location of the TBF during the time it was visible. This was the case at 08:21 UT, that is, during the first GRIS scan. This map will be analyzed in the following subsections. The contour of this TBF is shown in magenta in Fig.~\ref{inv_maps}. The positions of all other TBFs is shown as orange dotted lines in the GRIS continuum map to keep the visibility. The other TBFs, however, mostly appear at a different time than the time for which the TiO-band image is shown in Fig.~\ref{inv_maps}. Therefore, not all of them are visible on the upper left image obtained with the TiO-band filter. Green contours between umbra and penumbra in Fig.~\ref{inv_maps} represent contours produced from TiO-band images and brown contours represent contours produced from GRIS slit-reconstructed continuum intensity maps.

\subsection{Comparison between the penumbral region and the region with TBFs}
\label{compregions}
In this section, we investigate whether the visually apparent similarity between the deep photospheric magnetic topology on the western side of the sunspot (with penumbra) and on the eastern side of the sunspot (without penumbra) is reflected in the result of a quantitative analysis or not. We chose two rectangular regions of interest with equal height and width on either side of the sunspot. They are depicted with dashed boxes A and B in Fig.~\ref{inv_maps}. The widths were chosen such that the full penumbra is covered in box B, while the left-right locations were chosen such that they both border on the umbra on the spot-facing side. The heights and the up-down location of the boxes are also the same and were chosen such that a) no dark umbral-like structure at the bottom of region A is included, and b) the spurious line of discontinuity (see Sect.~\ref{grazam}) in the VFISV inclination map above region B is excluded.

\begin{figure*}[h]
\centering
   \includegraphics[width=17cm]{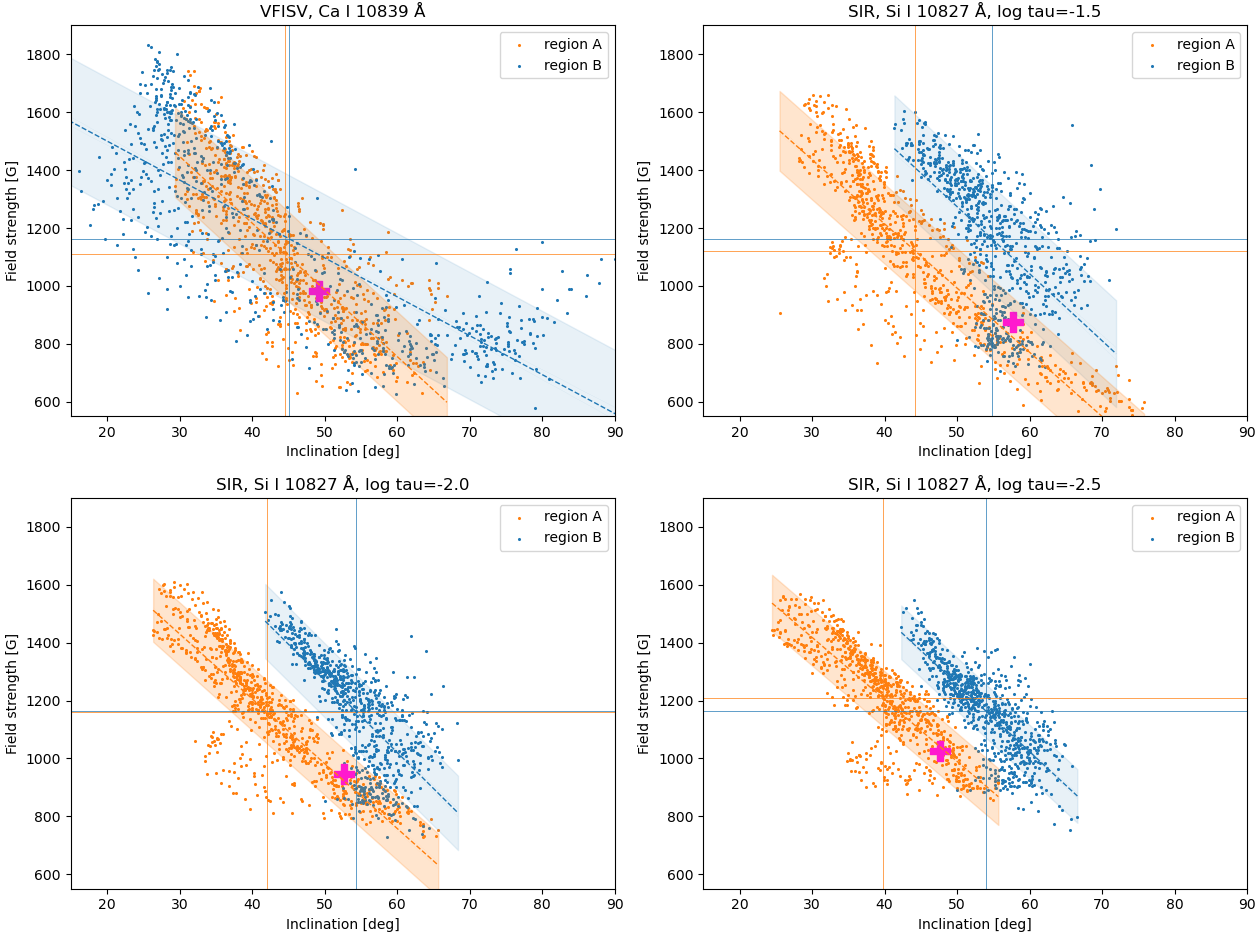}
     \caption{Scatter plot of magnetic field strength versus inclination for region A (with TBFs) in orange and Box B (penumbra) in blue. Averages over the respective regions are depicted as vertical and horizontal solid lines. The shaded areas depict the standard deviation from linear fits, which are shown as dashed lines. Magenta crosses show the spatial average inside the TBF contour plotted in magenta in Fig.~\ref{inv_maps}}
     \label{scatterplots}
\end{figure*}

\noindent In Fig.~\ref{scatterplots}, the magnetic field strength and inclination of all pixels in region A and region B are shown in scatter plots. This was done separately for the VFISV inversion map showing the deep photosphere and for several maps from the SIR inversion at different optical depth values. The scatter plots were produced from maps without an alignment of the GRIS to the BBI data to avoid spatial interpolations between pixels. The number of pixels inside the boxes was 750. In addition, the average inclination and field strength of the pixels inside the regions is given in Table \ref{avg_table}, together with the respective standard deviations.

\noindent All of the distributions follow a trend of high inclination values (more horizontal fields) being associated with low field strength values and vice versa. In a standard penumbra, this behavior would be expected from the spine (low inclination, high field strength) and intraspine (high inclination, low field strength) structure. In the deepest atmospheric layer (VFISV inversion map), the distributions from region A (in orange) and region B (in blue) largely overlap. This is also visible from the overlap of the areas corresponding to the standard deviation around a linear fit of distribution A and B. In addition, the average inclination value from region A and region B (depicted by vertical lines) differ only by \SI{0.5}{\degree}. Interestingly, this behavior changes in higher atmospheric layers: In the scatter plots from the SIR maps at atmospheric height $\log \tau = -1.5$, the population from region A is clearly separated from the population of region B. The average inclination value differs by \SI{10.5}{\degree}. When going to higher layers at $\log \tau = -2.0$ and $\log \tau = -2.5$, the differences become greater (\SI{12.3}{\degree} and \SI{14.2}{\degree}, respectively) and the shaded areas from the linear fits and their standard deviations do not overlap anymore. In summary, the magnetic topology of region A is similar to the one of region B in low atmospheric heights and becomes increasingly different towards higher layers. \\
As the abundance of inclined magnetic fields in the deep photosphere of region A (without penumbra) is unexpected, we checked whether these inclined fields persists or exists only for a short time. The spatial average of the inclination from the VFISV maps was calculated in region A also for the second and third GRIS scan, which were partially recorded under worse seeing conditions. The values \SI{44.9}{\degree} and \SI{43.3}{\degree} were obtained, which are close to the value \SI{44.6}{\degree} of the first GRIS scan. The inclined fields therefore exist for a minimum of approximately one hour. \\
In addition, we note that in the scatter plot of the deep photospheric VFISV inversions, only region B shows a `tail' of pixels with low field strength and high inclination (close to horizontal field). Studying the inversion map from Fig.~\ref{inv_maps}, we identify this tail with pixels at the outer penumbral boundary (see also the following Sect.~\ref{discussion}, point 2).

\begin{table}
\caption{Magnetic field inclination (`inc') with respect to the local vertical direction and magnetic field strength ('B'), averaged over region A (with TBFs) and region B (with penumbra). The abbreviation `std' stands for the respective standard deviation. Region A and region B each consisted of 750 pixels.}
\label{avg_table}
\centering 
\begin{tabular}{c c c | c c c c}
\hline
\hline
% use \rule to insert a box of certain height into empty cell. This increases the height of the first row.
 \rule{0pt}{2.5ex} & log $\tau$ & region & $\overline{\mbox{inc}}$ & std & $\overline{\mbox{B}}$ & std  \\
\hline 
VFISV & - & A & \SI{44.6}{\degree} & \SI{8.3}{\degree} & \SI{1111}{\gauss} & \SI{245}{\gauss} \\
VFISV & - & B & \SI{45.1}{\degree} & \SI{18.3}{\degree} & \SI{1163}{\gauss} & \SI{329}{\gauss} \\
\hline
SIR & -1.5 & A & \SI{44.2}{\degree} & \SI{10.6}{\degree} & \SI{1120}{\gauss} & \SI{271}{\gauss} \\
SIR & -1.5 & B & \SI{54.8}{\degree} & \SI{5.7}{\degree} & \SI{1164}{\gauss} & \SI{225}{\gauss} \\
\hline
SIR & -2.0 & A & \SI{42.0}{\degree} & \SI{8.3}{\degree} & \SI{1162}{\gauss} & \SI{215}{\gauss} \\
SIR & -2.0 & B & \SI{54.3}{\degree} & \SI{5.3}{\degree} & \SI{1163}{\gauss} & \SI{183}{\gauss} \\
\hline
SIR & -2.5 & A & \SI{39.8}{\degree} & \SI{6.7}{\degree} & \SI{1209}{\gauss} & \SI{173}{\gauss} \\
SIR & -2.5 & B & \SI{54.0}{\degree} & \SI{4.9}{\degree} & \SI{1163}{\gauss} & \SI{147}{\gauss} \\
\hline
HAZEL & - & A$^*$ & \SI{24.2}{\degree} & \SI{6.7}{\degree} & \SI{1197}{\gauss} & \SI{310}{\gauss} \\
HAZEL & - & B & \SI{68.0}{\degree} & \SI{25.1}{\degree} & \SI{526}{\gauss} & \SI{126}{\gauss} \\
\hline
\hline
\end{tabular}
\tablefoot{
\tablefoottext{*}{Only 460 pixels are used (see Sect.~\ref{hazel_inv})}
}
 \end{table}

\begin{figure}[h!]
\centering
	\resizebox{\hsize}{!}{\includegraphics{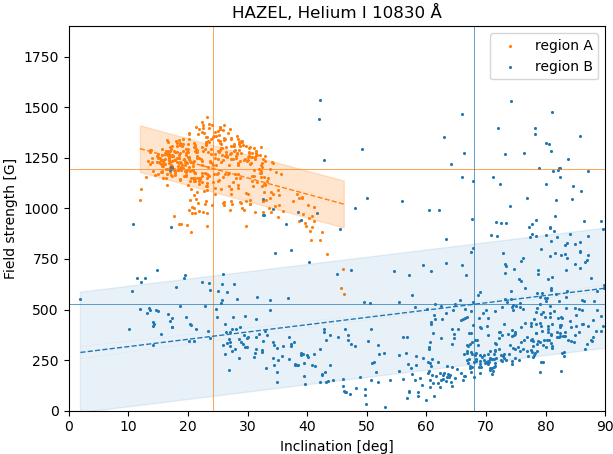}}
     \caption{Scatter plot of magnetic field strength and inclination from HAZEL inversion of the chromospheric \ion{He}{I}~10830~\SI{}{\angstrom}~triplet. For region B (penumbra), 750 pixels were used, but for region A (with TBFs) only 460 pixels could be used (see Sect.~\ref{hazel_inv}).} 
     \label{scatter_hazel}
\end{figure}

\noindent The magnetic field vector maps from the HAZEL inversion of the \ion{He}{I}~10830~\SI{}{\angstrom}~triplet allow for an analysis of the magnetic field in the high chromosphere, to which the other two lines are not sensitive to. As described in Sect.~\ref{hazel_inv}, some pixels had to be excluded from the analysis, which are located mostly on the East side of the sunspot. Region B was not affected, but in region A, 290 of the 750 pixels (\SI{39}{\percent}) had to be masked out. This limits the significance of our analysis, but we nevertheless analyzed the scatter plot shown in Fig.~\ref{scatter_hazel}. In region B, the distribution is generally more widespread than in the case of the VFISV and the SIR results. In region A, the distribution is more narrow than in region B. There is no overlap of the shaded areas corresponding to the standard deviation around the linear fits. In region A, the number of data points is reduced by \SI{39}{\percent}, but it is still remarkable that there are almost no pixels of  region B in the low-inclination and high-field-strength regime of the plot. The average inclination values of region A is \SI{24}{\degree} and that of region B is \SI{68}{\degree}. The average magnetic field strength are \SI{1197}{\gauss} and \SI{526}{\gauss} for region A and region B, respectively. \\
\noindent We tested whether the reduction of the number of pixels in region A influenced the scatter plot analysis by artificially reducing the pixel number also in region B. The pixel mask of region A was flipped  horizontally and then applied to region B. This ensured that the same symmetry with respect to the umbra is given. The shape of the resulting test distribution qualitatively stayed the same and the average inclination and field strength values changed by less than \SI{5}{\degree} and \SI{50}{\gauss}, respectively. Even in the imaginary and physically meaningless case of all excluded pixels having a fully horizontal field (inclination of \SI{90}{\degree}) and zero field strength, the average inclination would be \SI{50}{\degree} and would, therefore, still be lower than the average of region B. The magnetic field strength of this imaginary case would be \SI{734}{\gauss} for region A, which would also still be higher than the average field strength in region B. \\ \noindent The HAZEL inversion results of the high chromosphere therefore also show that the magnetic field is more vertical in region A than in region B. In addition, magnetic field strength values are higher in region A than in region B. We also note that the line depression map (Fig.~\ref{hazel}) shows that region A is covered by a darkening, while region B is not.

\subsection{Velocity field}
In the LOS velocity maps of the deep photospheric VFISV inversions (Fig.~\ref{inv_maps}), the penumbra on the right side of the image is mostly blueshifted. As the disk center direction is pointing towards the top right of the images and the observations were done at a heliocentric angle of \SI{55}{\degree}, this corresponds to a horizontal flow direction towards the top-right, an up-flow, or a combination of both. The flow on the right side of the sunspot is therefore consistent with the standard outward-directed photospheric Evershed flow. On the left side of the sunspot (in and around region A), large patches of redshifts are present which cover the entire TBF (magenta contour), but also smaller patches of blueshifts. The predominant redshifts have smaller magnitude than the blueshifts in and around region B, but could also represent outward-directed flows. A comparison to the inclination map shows that the red velocity patches mostly correspond to regions with more horizontal fields. Close to the umbral boundary, a band of small blueshifts is present. Also the velocity pattern in and around region A is therefore, in principle, compatible with an outward-directed Evershed flow, including up-flows close to the umbra \citep[see, e.g.][]{2011LRSP....8....4B}. A discussion whether these flows represent an Evershed flow follows in Sect.~\ref{discussion}. We also note that the blueshifts in and close to region A are mostly confined to bright features seen in the TiO-band image. \\
In the higher atmospheric layers shown by the maps of the SIR inversions at $\log \tau = -2.0$, redshifts are predominant in the penumbra in and around region B, which is consistent with the start of the in-flows in the superpenumbra. In and around region A, both blueshifts and redshifts are present. \\
The velocity maps from the HAZEL inversion show large-scale structures, as  shown in Fig.~\ref{hazel}: Outside the dark region seen in the helium line depression map, blueshifts are predominant on the East side and redshifts are predominant on the West side. This would also be consistent with the in-flow in the superpenumbra. Inside the dark region seen in the helium line depression map, both blueshifts and redshifts are present and the structure seem to be more complex.

\subsection{Atmospheric parameters at the TBF}
As described in Sect.~\ref{topology}, one contour of one TBF (seen in the TiO-band images) can be co-temporally analyzed in the GRIS inversion results shown in Fig.~\ref{inv_maps}. The contour covers 447 pixels in the BBI images and therefore also in the co-aligned and interpolated inversion maps. This corresponds to roughly 13 pixels in maps with the original GRIS pixel size (see image scales of the instruments in Sect.~\ref{data}). Although the TBF is located in the region in which filamentation is seen in the VFISV maps and the filamentation is parallel to the TBFs, we cannot unambiguously connect the TBF to one specific filament in the magnetograms. However, all TBFs that we identified were located in the region showing filamentation in inversion maps (in and around region A) and the orientations of TBFs and the filaments in the inversion maps were mostly parallel (see orange dotted lines in Fig.~\ref{inv_maps}). In order to further characterize the TBFs, we calculated average atmospheric parameters within the magenta contour outlining the TBF in Fig.~\ref{inv_maps}. 

\noindent The resulting numbers for the inclination and the field strength of the magnetic field, the velocity and the temperature are shown in Table~\ref{tbf_table}. These averages were not calculated for the chromospheric results from the HAZEL inversions because we clearly relate the observed TBFs to photospheric structures. The average values for the magnetic field parameters can also be read from the position of the magenta crosses in Fig.~\ref{scatterplots}. The crosses are located within the standard deviation band of the respective contributions, but towards the bottom right end of the distribution (more horizontal field and low field strength). This is also where the intraspine pixels of a penumbra would be located. In the deep photospheric VFISV velocity maps, redshifts are visible for the entire extent of the TBF, while the SIR velocity from $\log \tau = -2.5$ mostly shows blueshifts. We emphasize that the GRIS data do not resolve the TBFs and therefore can only describe the atmospheric surroundings. Especially, field strengths, inclinations and temperatures inside the TBFs can be higher than the values retrieved from the GRIS inversions.

\begin{table}
\caption{Upper part: Magnetic field inclination (`inc') with respect to the local vertical direction and magnetic field strength (`B'), averaged over the TBF shown by the magenta contour in Fig.~\ref{inv_maps}. Lower part: Same for the LOS-velocity (`vlos') and temperature (`T'). The abbreviation `std' stands for the respective standard deviation.}
\label{tbf_table}
\centering 
\begin{tabular}{c c | c c c c}
\hline
\hline
\rule{0pt}{2.5ex} & log $\tau$ & $\overline{\mbox{inc}}$ & std & $\overline{\mbox{B}}$ & std \\
\hline
VFISV & -  & \SI{48.9}{\degree} & \SI{4.9}{\degree} & \SI{990}{\gauss} & \SI{122}{\gauss} \\
SIR & -1.5  & \SI{57.2}{\degree} & \SI{13.3}{\degree} & \SI{887}{\gauss} & \SI{178}{\gauss} \\
SIR & -2.0  & \SI{52.3}{\degree} & \SI{9.5}{\degree} & \SI{955}{\gauss} & \SI{149}{\gauss} \\
SIR & -2.5  & \SI{47.4}{\degree} & \SI{5.8}{\degree} & \SI{1031}{\gauss} & \SI{129}{\gauss} \\
\hline
& & & & &  \\
 & log $\tau$ & $\overline{\mbox{vlos}}$ & std & $\overline{\mbox{T}}$ & std \\
 \hline
 VFISV & -  & \SI[per-mode=symbol]{+469}{\meter \per \second} & \SI[per-mode=symbol]{143}{\meter \per \second} & - & - \\
SIR & -1.5  & \SI[per-mode=symbol]{-583}{\meter \per \second} & \SI[per-mode=symbol]{422}{\meter \per \second} & \SI{4766}{\kelvin} & \SI{60}{\kelvin} \\
SIR & -2.0  & \SI[per-mode=symbol]{-853}{\meter \per \second} & \SI[per-mode=symbol]{542}{\meter \per \second} & \SI{4550}{\kelvin} & \SI{145}{\kelvin} \\
SIR & -2.5  & \SI[per-mode=symbol]{-687}{\meter \per \second} & \SI[per-mode=symbol]{537}{\meter \per \second} & \SI{4254}{\kelvin} & \SI{257}{\kelvin} \\
\hline
\hline
\end{tabular}

 \end{table}

\section{Discussion and Conclusion}
\label{discussion}
We present observations of a sunspot that developed a penumbra, except in the region in which the chromospheric magnetic field shows high field strength values and is close to vertically directed, which we interpret as a chromospheric disturbance. We find that:

\begin{enumerate}
\item In intensity, no penumbral filaments are seen on the eastern side of the sunspot (region A) for both high (BBI) and lower (GRIS) spatial resolution data. High-resolution images, however, show very thin bright filaments. We surmise that they are highly inclined magnetic flux tubes that are emerging. This is supported by the GRIS inversion results: They show that for the one examined TBF, the inclination and field strength are similar to those in the intraspines of a penumbra. HMI data show no significant sign of net flux emergence, but flux re-shuffling is still possible. Flux can emerge in one location and get advected by overturning convection in a different location \citep[see, e.g.,][]{2011ApJ...729....5R}. In addition, the spatial resolution of \SI{1}{\arcsec} limits the possibilities to observe smaller structures with HMI data.

\item In region A, the magnetic field parameters in deep photospheric layers reveal a magnetic topology that is similar to the classical penumbral structure seen on the opposite side of the spot (region B): The analysis of the top left scatter plot in Fig.~\ref{scatterplots} shows an overlap between the population of region A and the population of region B. In addition, the average inclination values of the two regions are almost identical, which shows that inclined magnetic fields are equally abundant. A visual inspection of the magnetic maps also show a spine and intraspine-like pattern on both sides. One difference between the regions, visible from the top left scatter plot in Fig. \ref{scatterplots}, is that region B shows a `tail' of pixels with low field strength and high inclination values, which corresponds to pixels at the outer edge of the penumbra. Such highly inclined magnetic fields at the outer penumbral boundary have already been observed for stable penumbrae \citep[e.g., by][]{2011LRSP....8....4B}. Intensity images clearly show that there are no penumbral filaments, but only TBFs in region A. From the deep photospheric magnetic and velocity maps and without looking at intensity images, one would expect a penumbra to be present. This is not the case. Thus, with this data set, we are showing that it is not possible to reliably tell whether a penumbra is present by looking at the magnetic topology in the deep photospheric layers alone. This is a surprising finding that, to our knowledge, has not been reported before.

\item The flow pattern in the deep photosphere would fit to an Evershed flow in region A. It is not possible to disentangle between out-flows and up-flows at this heliocentric angle, but since the magnetic field is inclined in this region, a flow with a significant horizontal component is most likely present. An alternative scenario for the dynamics observed in region A would be an inflow-outflow pattern similar to that reported by \cite{2019ASPC..526..261B} preceding penumbra formation.

\item Towards higher layers (SIR and HAZEL inversions), the scatter plots in Fig. \ref{scatterplots} and Fig. \ref{scatter_hazel} show differences between region A and region B, that increase with atmospheric height. Magnetic fields still have similar field strengths in the SIR inversions, but in region A, fields are more vertical than they are in region B. The almost vertical fields in the high chromosphere (average inclination value of \SI{24}{\degree}) of region A are in contrast to what one would expect from a classical chromospheric canopy: \citet[][Fig. 10d and 11d ]{2017A&A...604A..98J} reported values between \SI{40}{\degree} and \SI{80}{\degree}, which is in line with the average value of \SI{68}{\degree} we measured in region B showing a regular penumbra. The magnetic field strength values in the high chromosphere are also higher in region A (\SI{1197}{\gauss}) than in region B (\SI{526}{\gauss}). We interpret the strong vertical field in region A as a disturbance of the chromospheric canopy. This disturbance is also seen in the velocity maps and in helium line depression maps: Between the photosphere and the chromosphere (SIR and HAZEL inversion results), Fig.~\ref{inv_maps} shows a transition from the photospheric Evershed flow (outward-directed) to a superpenumbral inverse Evershed flow (inward-directed). Fig.~\ref{hazel} shows, however that the chromospheric inverse Evershed flow is disturbed in the region covered by the dark patch seen in the helium line depression map (eastern side of the FOV). In the remaining region of the sunspot, the inverse Evershed flow is seen as an in-flow pattern, like in other observations \citep[e.g.][Fig. 5d]{2017A&A...604A..98J}. The dark patch seen in the helium line depression map could represent magnetic loops that connect to the opposite polarity of the AR. This hypothesis is supported by the almost vertically directed (inclination: \SI{24}{\degree}) chromospheric magnetic fields: The magnetic fields are steep enough to represent footpoints of a loop  extending over a large distance.
\end{enumerate}

\noindent The sunspot we observed developed a penumbra except on the side where the chromospheric canopy is disturbed, although the ``photospheric ingredient'' for penumbra formation (deep photospheric magnetic fields of penumbra-like topology) are present on both sides. This suggests that an undisturbed chromospheric canopy is needed for the formation of a penumbra. This is, to our knowledge, the first evidence of the connectivity between the photosphere and chromosphere playing a role in penumbra formation based on spectropolarimetric data from both the photosphere and the chromosphere. We note that four days after our observations, HMI data (see Fig.~\ref{hmi_cont}) shows that a penumbra has developed in a large, but not the full part of the region that showed a chromospheric disturbance in our data. We cannot tell whether the chromospheric disturbance is still present or not four days after our observations. We therefore restrict our finding only to the first days of penumbra formation.

\noindent The sunspot structure we describe in this work is partially similar to the sunspot structure analysed by \citet{2013ApJ...769L..18L}, in which a penumbra formed on the side of a spot facing the opposite polarity, where small magnetic elements are appearing and disappearing again. The crucial difference, however, is that in their case, the canopy was not interrupted and a penumbra could form. We also compare our observations to the observations reported by \citet{2021A&A...653A..93M}, where a partial penumbra formed on the side of a pore in which magnetic field extrapolations based on HMI data show a canopy-like structure. The main difference to the present work is that in our case, inclined photospheric fields are present also in the region where no penumbra formed. The lack of  horizontal flux can therefore be excluded as a reason for the penumbra not to form. \\

\noindent Our results also have implications on the discussion of different penumbra formation scenarios: We demonstrate that no inclined chromospheric fields are needed for the stable (for a minimum of one hour) development of inclined photospheric fields: In region A, showing no penumbra, but TBFs, spatial averages of roughly \SI{45}{\degree} have been determined with suitable field  strength (\SI{1111}{\gauss}) in the deep photosphere (see Sect.~\ref{compregions}), which is similar to the average values measured in the penumbral region B. The HAZEL inversions show that in this region A, the high chromospheric magnetic fields are close to vertical with a spatial average of \SI{24}{\degree}. We therefore assume that the photospheric inclined fields origin from below the solar surface, questioning the penumbra formation scenario in which inclined field lines fall from the chromosphere onto the photosphere \citep{1992ApJ...388..211W,2012ApJ...747L..18S,2013ApJ...771L...3R,2016ApJ...825...75M}.
Our results instead support the bottom-up scenario of \citet{1998ApJ...507..454L}, \citet{2013ApJ...769L..18L}, \citet{2014ApJ...787...57Z}, \citet{2018ApJ...857...21L} and \citet{2019ApJ...886..149L}. \\
\begin{figure}[h]
\centering
	\includegraphics[width=7.5cm]{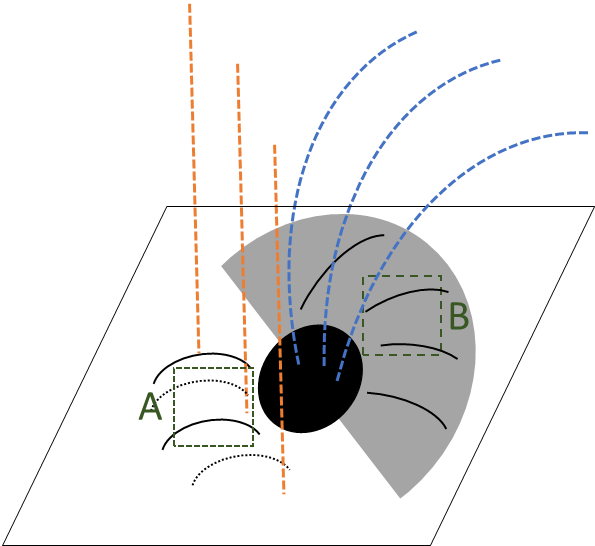}
     \caption{Schematic representation of the magnetic field configuration we propose for the present sunspot. Field lines observed in the chromosphere are shown in orange- and blue-dashed lines. Photospheric field lines are shown in solid and dotted black lines.} 
     \label{sketch}
\end{figure}

\noindent We propose the following scenario for this sunspot and refer to Fig.~\ref{sketch} for a sketch: Inclined fields in the penumbral chromospheric canopy of region B (blue-dashed lines) block the underlying horizontally aligned photospheric field lines from further rising into higher layers and a stable penumbra establishes. The intraspines of the penumbra are shown as solid black lines in the gray penumbral region in Fig.~\ref{sketch}. In region A, however, the canopy is disrupted, magnetic fields are more vertical (orange-dashed lines) and the horizontal field lines appearing in deep photospheric layers cannot be prevented from further rising. No penumbra can form. We identify the TBFs as examples of such flux ropes (shown as black lines in Fig.~\ref{sketch}) that would form the penumbra if a canopy with inclined magnetic fields was present above. If this is not the case, the flux ropes dissolve in the measured lifetime of roughly \SI{4}{\minute} or rise to higher layers instead of becoming part of a stable penumbra. This decaying process is represented in Fig.~\ref{sketch} as some black lines being shown as dotted lines. \\ 

\noindent The assumption that a rising magnetic flux tube is blocked by an overlying chromospheric magnetic field can be explained in ideal MHD: At the lower boundary of a horizontally aligned chromospheric canopy, the locally increasing background field exerts an expelling force on the rising flux \citep[see,][Equation 20]{1998A&A...337..897S}. In case the background field is close to vertical, one expects the background field to decrease with height. This would result in an upward pointing force further supporting the rise of a magnetic flux tube. \\

\noindent The exact nature of the chromospheric disturbance is still an open question, partially also because it extends beyond our FOV. More data, especially of the chromosphere, would be needed to further characterize such disturbances and investigate whether they are always connected to a missing penumbra or not.

  %==============================================
  %==============================================

\begin{acknowledgements} 

The 1.5-meter GREGOR solar telescope was built by a German consortium under the leadership of the Leibniz-Institute for Solar Physics (KIS) in Freiburg with the Leibniz Institute for Astrophysics Potsdam, the Institute for Astrophysics Göttingen, and the Max Planck Institute for Solar System Research in Göttingen as partners, and with contributions by the Instituto de Astrofísica de Canarias and the Astronomical Institute of the Academy of Sciences of the Czech Republic. The redesign of the GREGOR AO and instrument distribution optics was carried out by KIS whose technical staff is gratefully acknowledged. The GRIS instrument was developed thanks to the support by the Spanish Ministry of Economy and Competitiveness through the project AYA2010-18029 (Solar Magnetism and Astrophysical Spectropolarimetry). SJGM is grateful for the support of the European ResearchCouncil through the grant ERC- 2017-CoG771310-PI2FA, the Spanish Ministry of Science, Innovation and Universities (MCIU) through the grant PGC2018- 095832-B-I00, and by the project VEGA 2/0048/20. LK is supported by a SNSF PRIMA grant. CK acknowledges funding received from the European Union's Horizon 2020 research and innovation programme under the Marie Sk\l{}odowska-Curie grant agreement No 895955.
This research used version 1.1.3 of the SunPy open source software package \citep{sunpy_community2020}. 
\\ We thank Rolf Schlichenmaier for creating a first version of the reconstructed BBI data using the KISIP code which led to the identification of the thin bright filaments investigated in this paper, his useful suggestions on the velocity calibration and for fruitful discussions on several aspects of the paper. 
We are very thankful to Svetlana Berdyugina for providing the TiO-band filter optimized for the BBI observations, carrying out the GREGOR observations together with the lead author, obtaining the new BBI calibration data, collaborating on the BBI data pipeline and assuring the correct contrast of the thin bright filaments investigated in this paper, as well as for her advice on the statistical analysis of the inversion results and helpful comments on the manuscript.
We also thank Andrés Asensio Ramos for his help with the HAZEL inversion. In addition, we thank Oskar Steiner for a constructive feedback on the manuscript.

\\ Author contribution statement: PL designed and directed the research project. PL conducted the GREGOR observations. LK wrote the BBI calibration pipeline, revealed a calibration problem and provided essential guidance for the final calibration and reconstruction of the BBI data. PL discovered and characterized the 'thin bright filaments'. PL analyzed the HMI data. PL and LK calibrated the GRIS data. CK prepared the GRIS data for the inversions, including the initial wavelength calibration. PL ran the VFISV inversions. CK ran the SIR inversions. SJGM ran the HAZEL inversions. SJGM, CK and LK helped PL applying the GRAZAM code to resolve the azimuthal disambiguation. PL, CK, SJGM and NBG analyzed and discussed all inversion results, including the adaptation of inversion parameters and an improved velocity calibration. PL analyzed statistical correlations of the magnetic field strength and inclination from different inversion results. PL wrote the first version of the manuscript and especially NBG also contributed to the interpretation of the results. All authors contributed to the final manuscript. LK led and executed, together with TB, the redesign of the GREGOR AO and instrument distribution optics before the Science verification phase (organized by LK), during which the data for this research project was acquired.

\end{acknowledgements}

\bibliographystyle{bibtex/aa} % style aa.bst
\bibliography{thinfilambib} % your references Yourfile.bib

% ----------------------------------------------------------------------
\begin{appendix}
% ----------------------------------------------------------------------
\section{HMI magnetograms}
\label{app_hmi}
\begin{figure}[h!]
\centering
   \includegraphics[width=8.8cm]{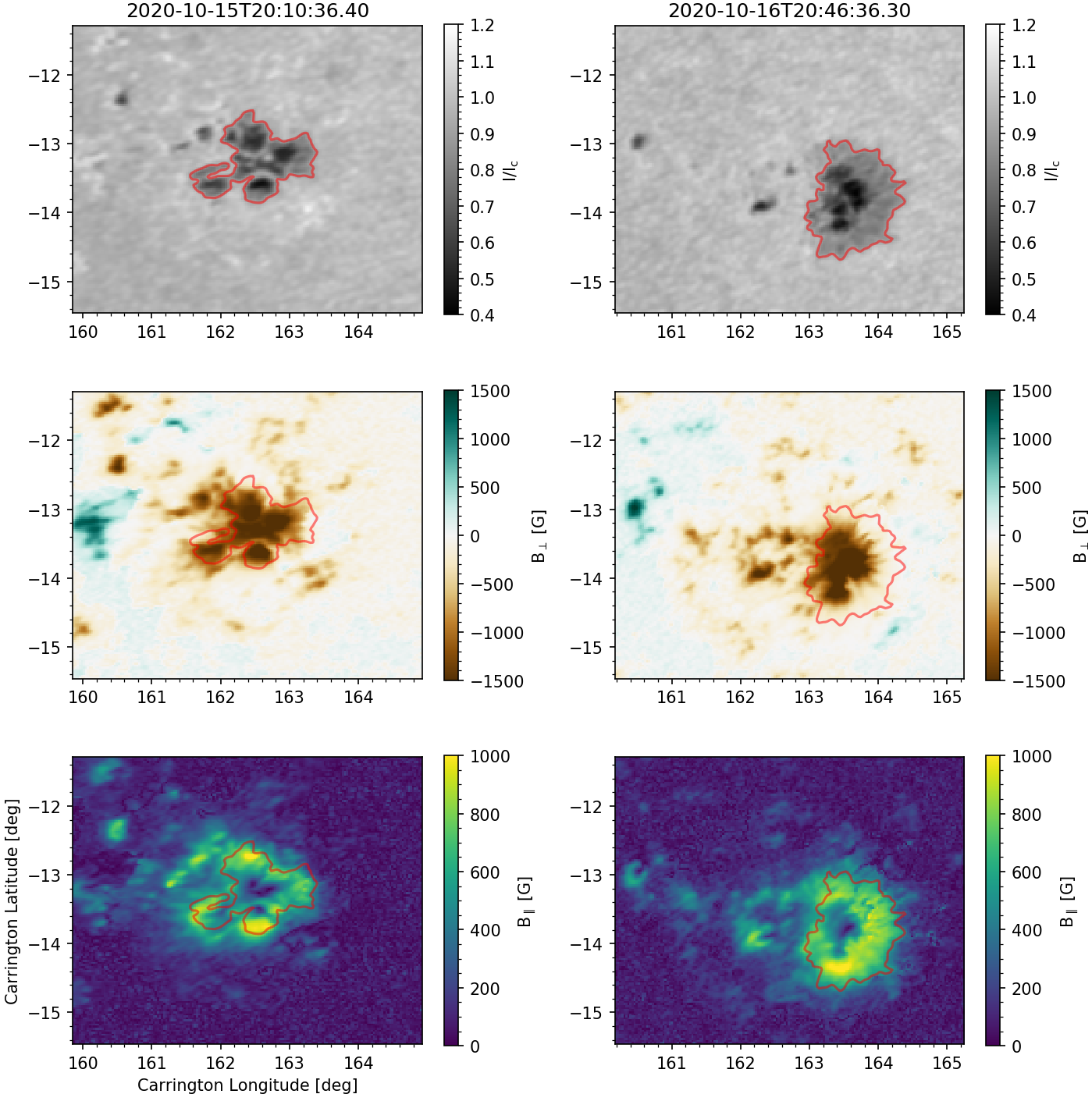}
     \caption{Maps showing the continuum intensity (top), the vertical component of the magnetic field (center) and the horizontal component of the magnetic field (bottom). The red line shows the intensity contour that was used to calculate the spot area.} 
     \label{hmi_cont_mag}
\end{figure}

The flux evolution shown in Fig.~\ref{fluxevol} was calculated based on the hmi.sharp\_cea\_720s data series. Maps of the continuum intensity, the vertical component of the magnetic field and the horizontal component of the magnetic field from two example time steps are shown in Fig.~\ref{hmi_cont_mag}. The intensity was normalized by a nearby quiet-sun average. The red line shows the contour at the intensity level of 0.9, that was used to calculate the spot area.

% ----------------------------------------------------------------------
\section{Lengths of additional TBFs}
\begin{figure}[h!]
\centering
    \includegraphics[width=7.5cm]{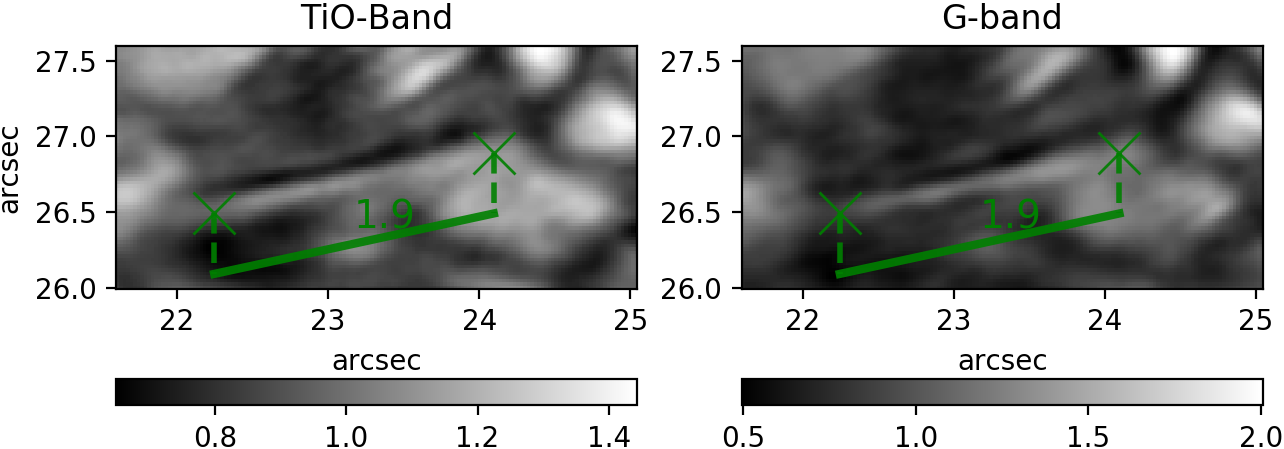}
    \includegraphics[width=7.5cm]{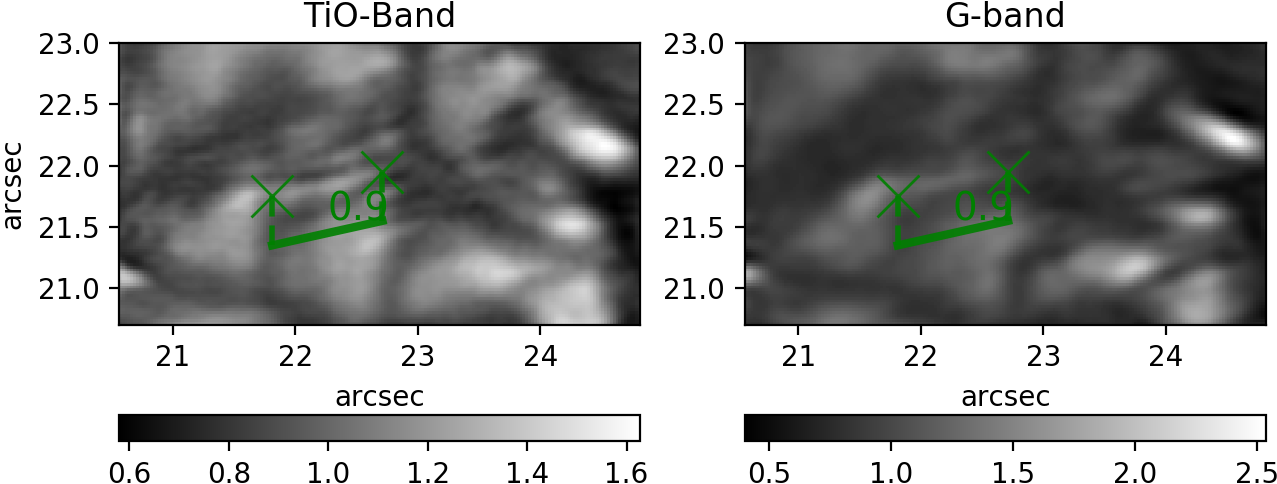}
    \includegraphics[width=7.5cm]{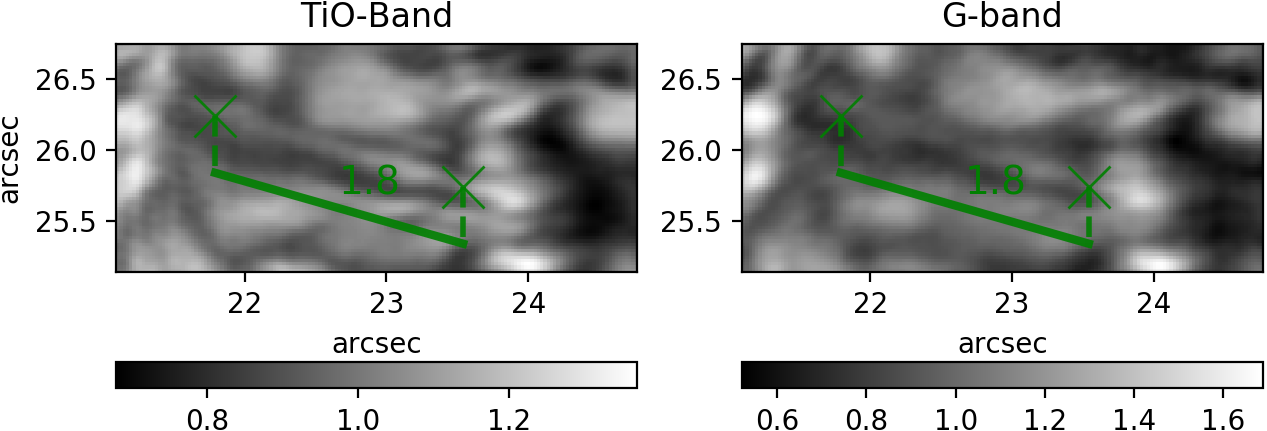}
    \includegraphics[width=7.5cm]{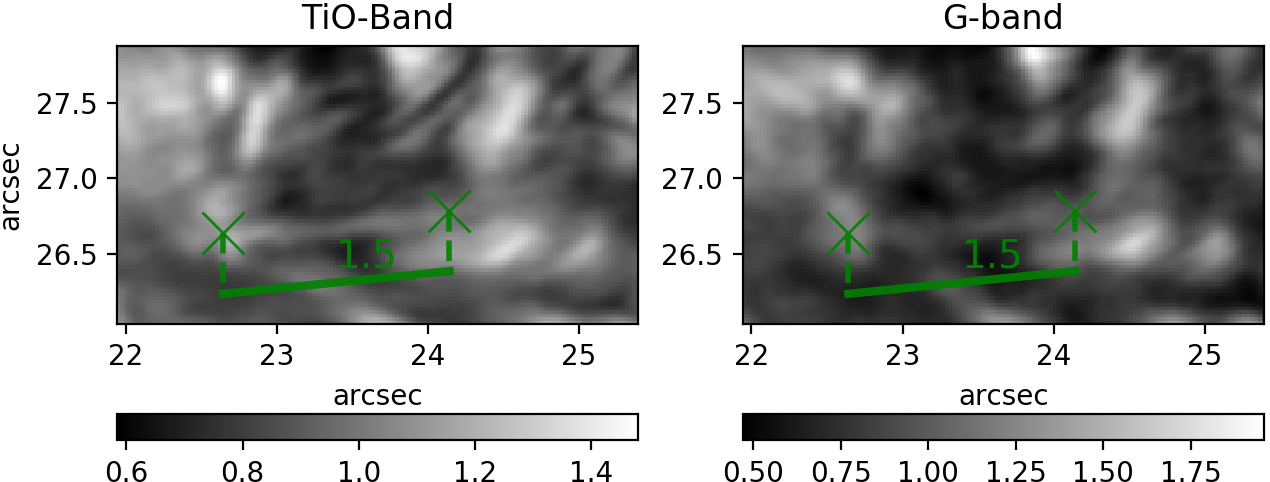}
    \includegraphics[width=7.5cm]{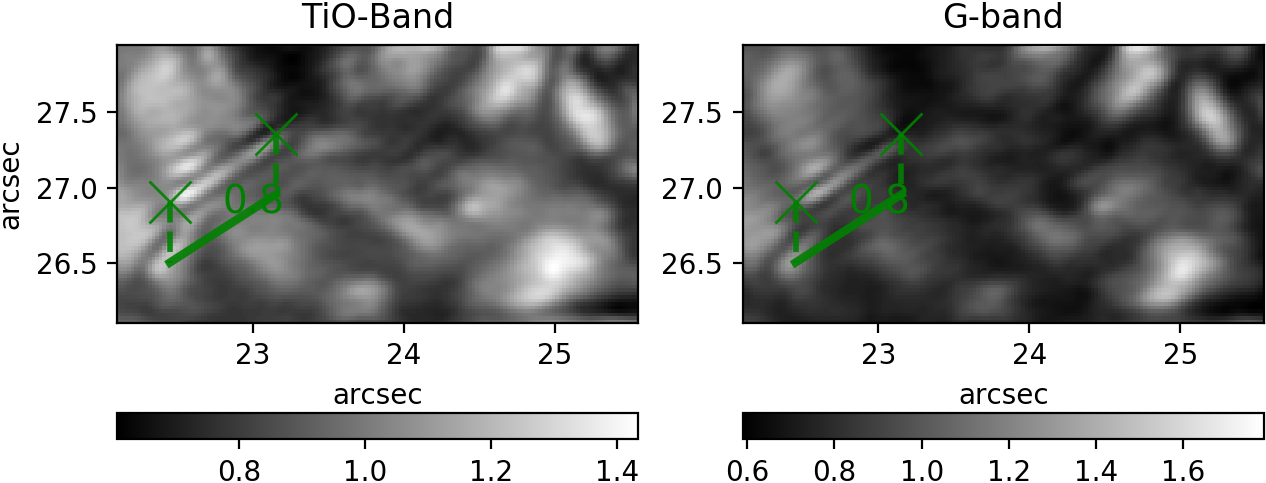}
    \includegraphics[width=7.5cm]{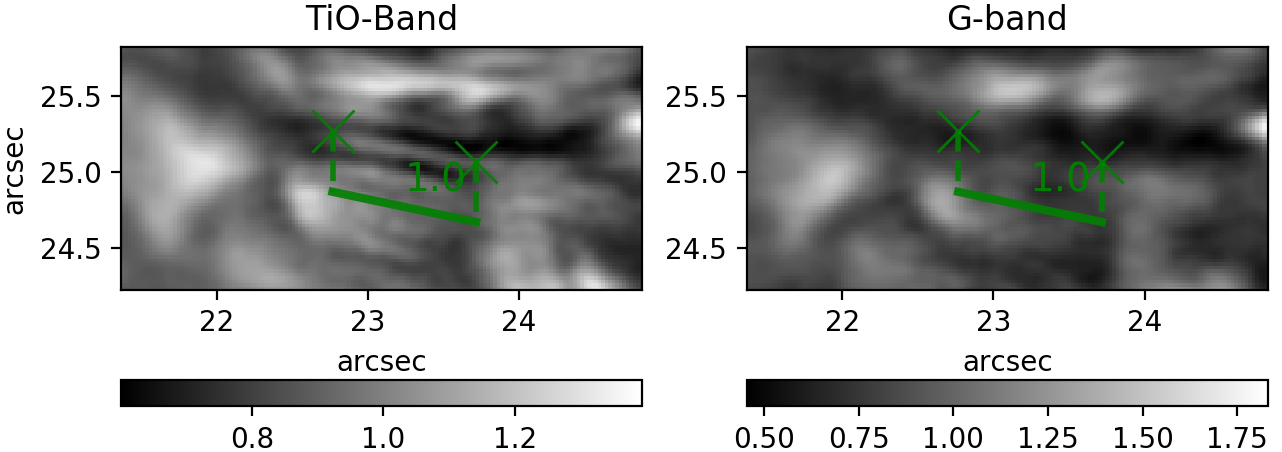}
    \includegraphics[width=7.5cm]{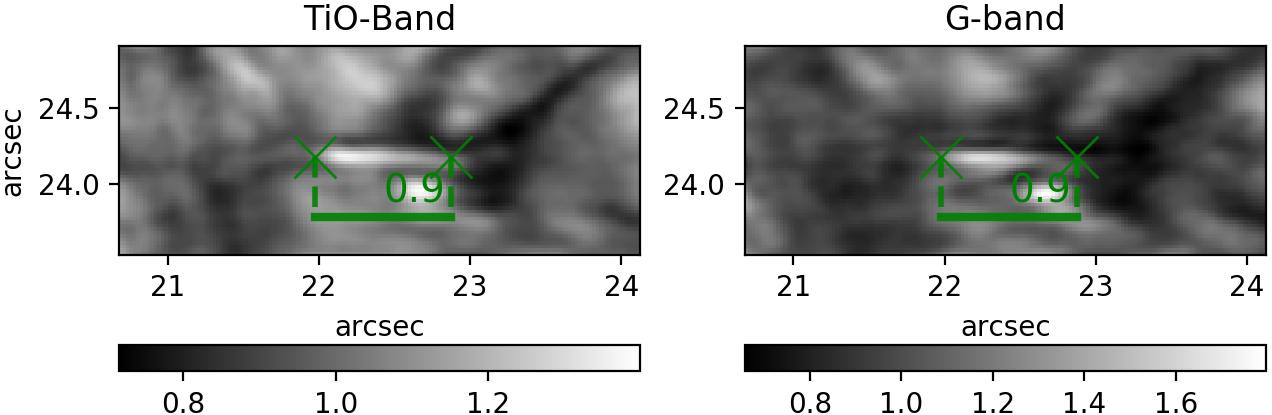}
    \includegraphics[width=7.5cm]{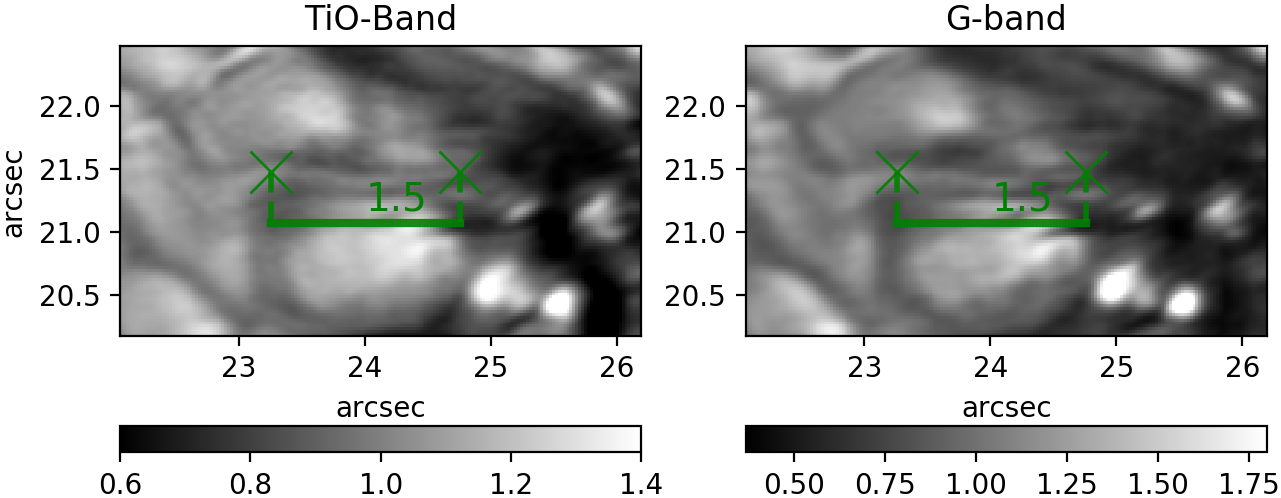}
    \caption{Images from BBI of TBFs obtained with the TiO-band filter (left) and the G-band filter (right), in addition to the one shown in Fig.~\ref{morph_example}. } 
    \label{all_tbfs}
\end{figure}

Figure~\ref{all_tbfs} shows BBI images of the TBFs we manually identified in addition to the TBF shown in Fig.~\ref{morph_example}. Both images obtained with the TiO-band filter and the G-band filter are shown. The TBFs have been observed at different times. The green lines rulers and numbers show the length measurement we have manually done.

\label{app1}

% ----------------------------------------------------------------------
\section{Inversion fit for different pixels}
\label{app_examspecs}
\begin{figure*}[h]
\centering
   \includegraphics[width=17cm]{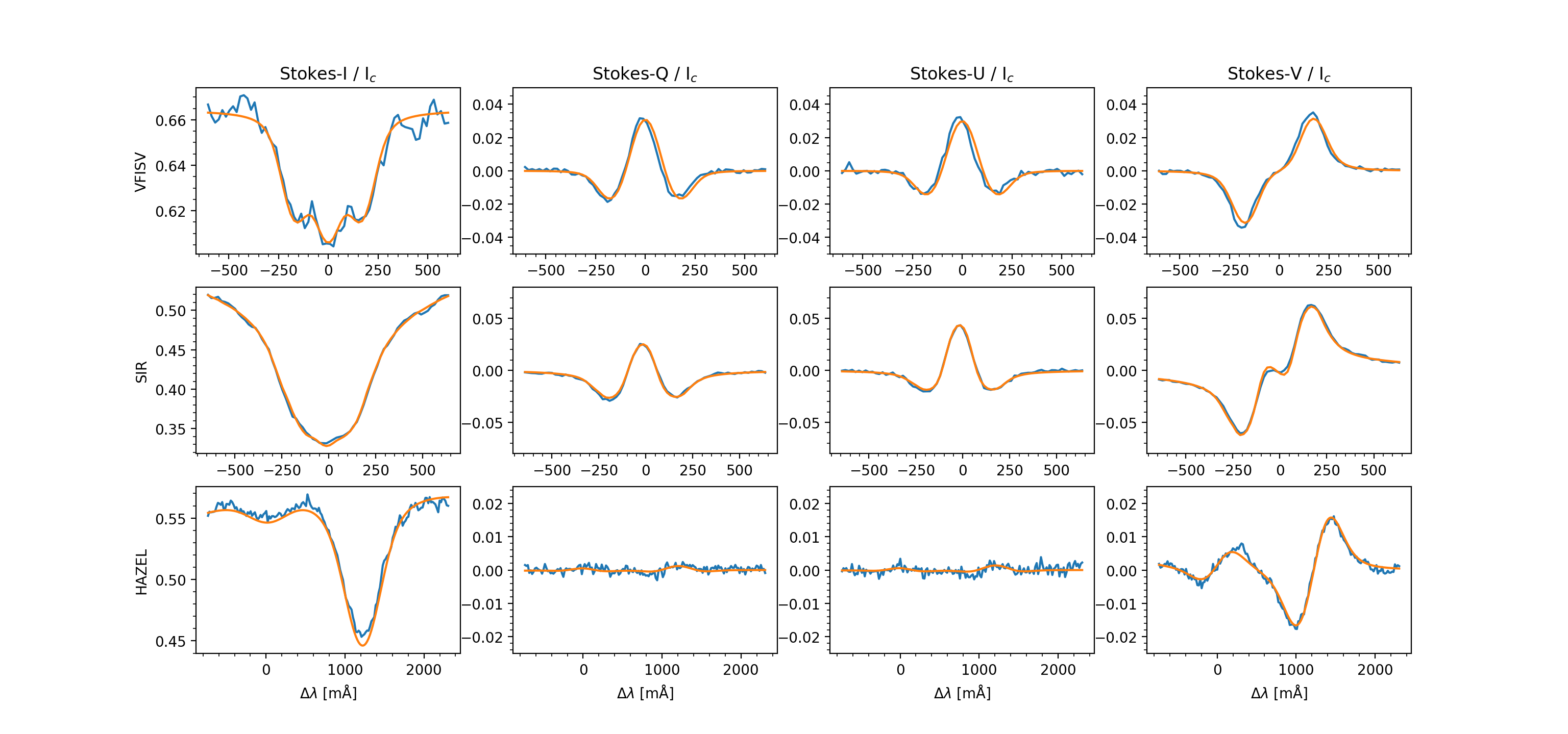}
     \caption{Spectral profiles (observations  in blue, inversion fit in orange) of a pixel from the umbra. The (x,y) coordinates of the location are (\SI{28.6}{\arcsec}, \SI{26.6}{\arcsec}). Top row: \ion{Ca}{I}~10839~\SI{}{\angstrom} line, central row: \ion{Si}{I}~10827~\SI{}{\angstrom} line, bottom row: \ion{He}{I}~10830~\SI{}{\angstrom} triplet. The Stokes-I profile of the \ion{Ca}{I}~10839~\SI{}{\angstrom} line shows Zeeman splitting features because of the large magnetic field in the deepest layer.} 
     \label{examspec_umbra}
\end{figure*}

\begin{figure*}[h]
\centering
   \includegraphics[width=17cm]{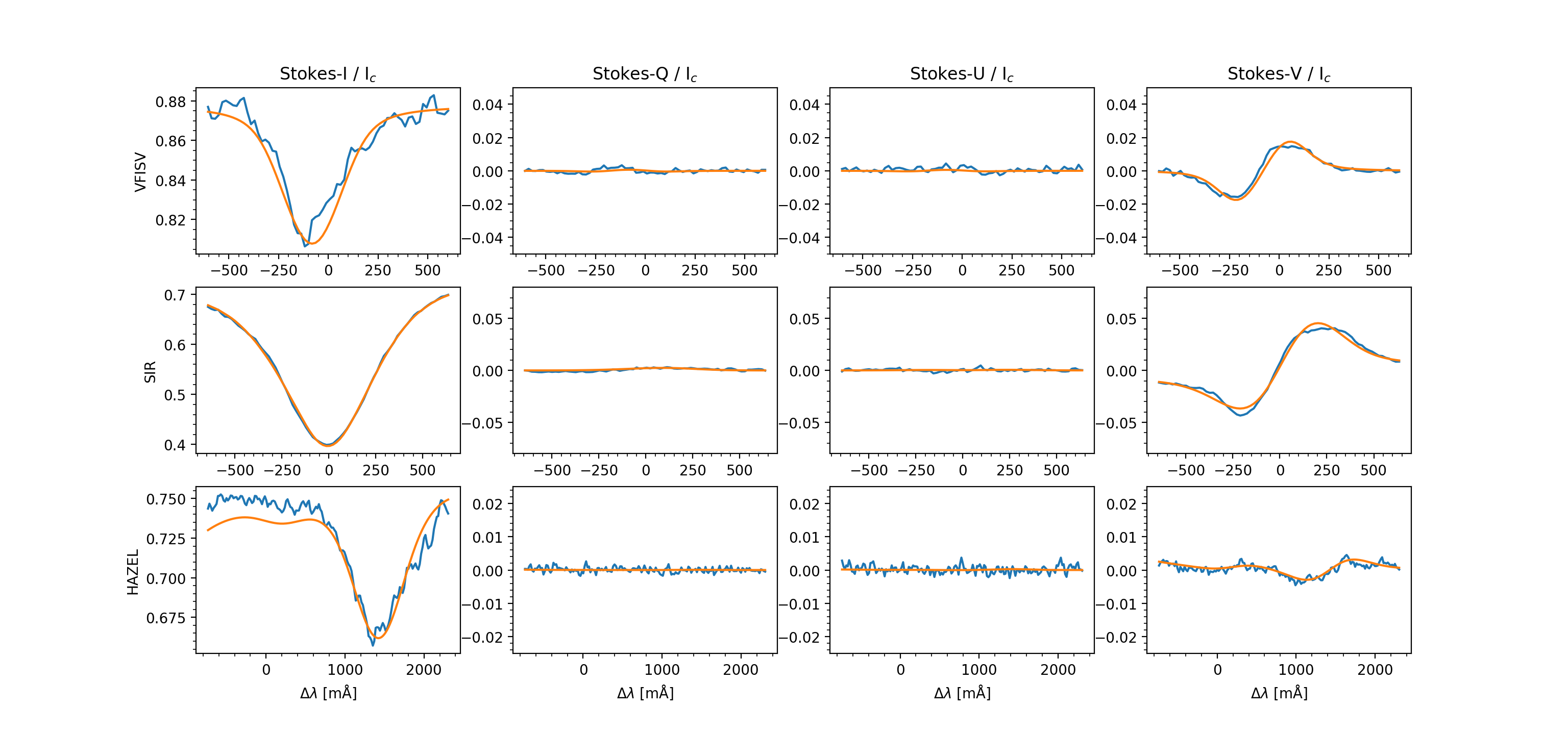}
     \caption{ Spectral profiles (observations  in blue, inversion fit in orange) of a pixel from the penumbra. The (x,y) coordinates of the location are (\SI{36.8}{\arcsec}, \SI{25.6}{\arcsec}). Top row: \ion{Ca}{I}~10839~\SI{}{\angstrom} line, central row: \ion{Si}{I}~10827~\SI{}{\angstrom} line, bottom row: \ion{He}{I}~10830~\SI{}{\angstrom} triplet.} 
     \label{examspec_penumbra}
\end{figure*}

\begin{figure*}[h]
\centering
   \includegraphics[width=17cm]{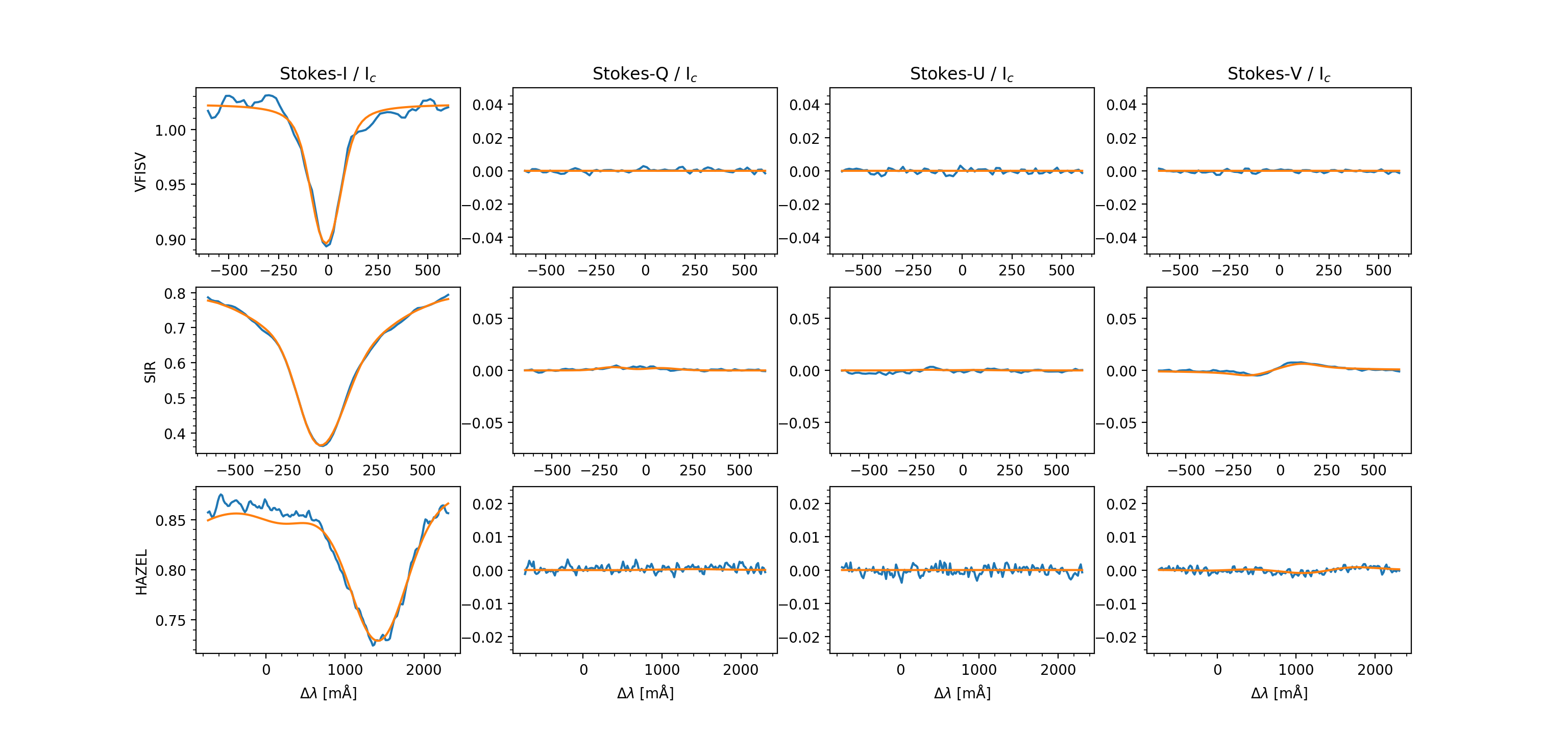}
     \caption{pectral profiles (observations  in blue, inversion fit in orange) of a pixel from the quiet sun. The (x,y) coordinates of the location are (\SI{52.7}{\arcsec}, \SI{23.9}{\arcsec}). Top row: \ion{Ca}{I}~10839~\SI{}{\angstrom} line, central row: \ion{Si}{I}~10827~\SI{}{\angstrom} line, bottom row: \ion{He}{I}~10830~\SI{}{\angstrom} triplet.} 
     \label{examspec_qs}
\end{figure*}

In Fig.~\ref{examspec_umbra}, Fig~\ref{examspec_penumbra}, and Fig~\ref{examspec_qs}, spectral profiles of pixel from the umbra, penumbra and the quiet sun are shown. The obserations are shown as blue lines and the respective inversion fit as orange lines. All spectra are normalized to the quiet-sun continuum value.

\end{appendix}

\end{document}